\documentclass[]{aa}

\usepackage[varg]{txfonts}
\usepackage[normalem]{ulem}
\usepackage{tikz}
\usetikzlibrary{shapes, arrows.meta, positioning}

\RequirePackage[]{xcolor}
\definecolor{AIPblue}      {RGB}{  0, 65,120}
\definecolor{AIPorange}    {RGB}{242,148,  0}
\definecolor{AIPlightgrey} {RGB}{230,230,230}
\definecolor{AIPdarkgrey}  {RGB}{ 80, 85, 90}
\definecolor{tikzgreen}    {RGB}{0,153,25}
\definecolor{tikzyellow}   {RGB}{255,255,77}

\newcommand\hpsstars{18} 
 
\newcommand\agreeingbase{193} 
\newcommand\agreeingfast{43} 
 
\newcommand\agreeingallbase{236}

\newcommand\disagreeingwb{38} 
\newcommand\binaryevo{21}
\newcommand\binarywd{68} 
\newcommand\mswidebinaries{283} 
\newcommand\alltwoperiodwidebinaries{292} 
\newcommand\mswidebinariesbase{274} 
\newcommand\allbinaries{372}

\newcommand\outlierbinaries{29}
\newcommand\wbslowbinaries{29} 
\newcommand\wbfastbinaries{18} 
\newcommand\wballbinaries{76}

\newcommand\brokenbinaries{131} 
 
\newcommand\alldismissed{196}

\newcommand\closeperiods{3086}
\newcommand\changedperiods{52} 
\newcommand\allevolved{89}

\newcommand\startingbinariesall{3124}
\newcommand\binariestwoperiods{210} 
\newcommand\evolvedwithtwoperiods{7} 
\newcommand\evolvedwithoneperiod{12} 
\newcommand\evolvedms{19} 
\newcommand\startsamplestars{24\,404} 
\newcommand\startsamplebinaries{12\,202} 
\newcommand\startsamplebinarieswperiods{3367} 
\newcommand\startsamplebinarieswonep{2783} 
\newcommand\startsamplebinariestotal{3075} 
\newcommand\msmsoneperiod{2703} 
\newcommand\startsampleperiodrhall{15\,278} 
\newcommand\startsampleperiodrh{1465} 

\newcommand\startsampleperiodmqall{9126} 
\newcommand\startsampleperiodmq{1869} 
 
\newcommand\alllitperiods{3334}

\newcommand\alladditonaltoall{229} 
\newcommand\alladditonaltorh{102} 
\newcommand\alladditonaltomq{127} 
\newcommand\allmissingtoall{65}

\newcommand\targetoffevolved{260} 
\newcommand\targetbothevolvedb{372}
\newcommand\targetbothevolveda{346}
\newcommand\hierarchicaloutlier{297}
\newcommand\binarieswgaiametalicity{231} 
\newcommand\binarieswticmetalicity{33} 

\newcommand{\class}[1]{{\it #1}}
\newcommand{\classS}{\class{S}}
\newcommand{\classF}{\class{F}}
\newcommand{\classC}{\class{C}}
\newcommand{\classE}{\class{E}}
\newcommand{\classW}{\class{W}}
\newcommand{\primary}{{\emph{L}}}
\newcommand{\secondary}{{\emph{T}}}

\defcitealias{2023AnA...672A.159G}{GBW23} 
\defcitealias{2021MNRAS.506.2269E}{EB21}
\defcitealias{2020AnA...635A..43R}{RH20}
\defcitealias{2014ApJS..211...24M}{MQ14}
\defcitealias{2016AJ....152..100L}{\texttt{Everest}}
\defcitealias{2022arXiv220800211G}{GDR3}
\defcitealias{2022yCat.4039....0P}{TIC}
\defcitealias{2017yCat.4034....0H}{EPIC}
\defcitealias{2009yCat.5133....0K}{KIC}
\defcitealias{2003yCat.2246....0C}{\textsc{2mass}}
\defcitealias{2012MNRAS.427..127B}{\texttt{PARSEC}}
\defcitealias{2000AnAS..143....9W}{\textsc{Simbad}}
\defcitealias{2023ApJ...947L...3B}{BPH23}

\usepackage[colorlinks=true,citecolor=AIPblue,linkcolor=AIPblue,urlcolor=AIPdarkgrey]{hyperref}

\begin{document}

\title{Wide binaries demonstrate the consistency of rotational evolution between open cluster and field stars%
    \thanks{The complete Table\,\ref{tab_period_sample} is only available in electronic form at the CDS via anonymous ftp to cdsarc.u-strasbg.fr (130.79.128.5) or via http://cdsweb.u-strasbg.fr/cgi-bin/qcat?J/A+A/675/A180}%
    }

\subtitle{}

\author{D. Gruner\inst{1,2}, S. A. Barnes\inst{1,3}, K. A. Janes\inst{4}}

\institute{
    Leibniz-Institute for Astrophysics Potsdam (AIP), An der Sternwarte 16, 14482 Potsdam, Germany
        \and 
    Institut für Physik und Astronomie, Universit\"at Potsdam, Karl-Liebknecht-Str. 24/25, 14476 Potsdam, Germany
        \and
    Space Science Institute, 4765 Walnut St STE B, Boulder, CO 80301, USA
        \and
    Astronomy Department, Boston University, 685 Commonwealth Avenue, Boston, MA 02215, USA
    }

\date{Received 4 April 2023 / Accepted 22 May 2023}

\abstract 
  {
    Gyrochronology enables the derivation of ages of late-type main sequence stars based on their rotation periods and a mass proxy, such as color. It has been explored in open clusters, but a connection to field stars has yet to be  successfully established.
  }
  {
    We explore the rotation rates of wide binaries, representing enlightening intermediaries between clusters and field stars, and their overlap with those of open cluster stars. 
  }
  {
    We investigated a recently created catalog of wide binaries, matched the cataloged binaries to observations by the \emph{Kepler} mission (and its \emph{K2} extension), validated or re-derived their rotation periods, identified \mswidebinaries{} systems where both stars are on the main sequence and have vetted rotation periods, and compared the systems with open cluster data.
  }
  {
    We find that the vast majority of these wide binaries (\agreeingallbase{}) line up directly along the curvilinear ribs defined by open clusters in color-period diagrams or along the equivalent interstitial gaps between successive open clusters. The parallelism in shape is remarkable. Twelve additional systems are clearly rotationally older. The deviant systems, a minority, are mostly demonstrably hierarchical. Furthermore, the position of the evolved component in the color-magnitude diagram for the additional wide binary systems that contain one is consistent with the main sequence component's rotational age.
  }
  {
    We conclude that wide binaries, despite their diversity, follow the same spindown relationship as observed in open clusters, and we find that rotation-based age estimates yield the same ages for both components in a wide binary. This suggests that cluster and field stars spin down in the same way and that gyrochronology can be applied to field stars to determine their ages, provided that they are sufficiently distant from any companions to be considered effectively single. The results also suggest that the impact of metallicity variations on the spindown is likely not to be a major concern.
  }

\titlerunning{Rotation of wide binaries}

\authorrunning{Gruner, Barnes \& Janes}


\maketitle

\section{Introduction} \label{sec_intro}

    Wide binary systems are unique in the sense that they combine aspects of both field stars and cluster stars in an advantageous way. With field stars, they share the aspects of diversity regarding their formation and composition, the continuous and vast range of their ages, and their scattered locations across the Galaxy. And just as with cluster stars, the shared origin and coevality of the components furnishes insights not gained from an isolated star, especially with respect to stellar evolution. Thus, they provide an opportunity to connect the behaviors of cluster and field stars by asking whether an assertion that is true for one component treated as a field star is also true for the other. This duality allows us to probe beyond the limited parameter space spanned by cluster stars while still benefiting from the intrinsic consistency of the individual components. In this paper, we demonstrate an underlying connection between the rotation and age behaviors of clusters and wide binaries, thereby strengthening the possibility of deriving ages through rotation (gyrochronology) for appropriately characterized cool field dwarfs.

    Reliable characterization of stars is essential to various fields of astronomy. From exoplanet hosts to stars in larger populations, understanding the star is key to understanding the system. However, a crucial parameter involved in such characterization is the rather elusive stellar age. But deriving reliable ages requires rather extensive data and cumbersome work, especially for late-type main sequence stars, which are the most frequent types of stars in the universe. The methods employed typically involve the observed correlation of one or more stellar parameters with age. However, classical stellar parameters either change very little over the course of billions of years (e.g., temperature, luminosity), can only be used reliably in young stars (e.g., Li abundance), or change systematically on long time scales while varying significantly on short time scales (e.g., activity). This imposes various limits on their applicability, especially during the long main sequence careers of cool stars, during which the classical parameters are almost constant. (For an extensive review of age dating methods see \cite{2010ARAnA..48..581S}.)

    Gyrochronology offers an age dating method that alleviates some of these problems, particularly on the main sequence. It exploits the observed decline in stellar rotation with age. The idea of this spindown goes back to the work of \cite{1972ApJ...171..565S} and scattered antecedents. Subsequently, it was found that the spindown is systematically mass dependent \citep[][see also \citealt{2007ApJ...669.1167B}]{2003ApJ...586..464B}. Numerous studies of stellar open clusters have since explored the spindown relationships further, including (but not limited to) the \object{Pleiades} \citep[125\,Myr, \cite{1987AnAS...67..483V}, revisited by][]{2016AJ....152..113R}, the \object{Hyades} \citep[650\,Myr, \cite{1987ApJ...321..459R}, revisited by][]{2019ApJ...879..100D}, \object{Praesepe} \citep[700\,Myr,][revisited by \citealt{2017ApJ...842...83D} and \citealt{2021ApJ...921..167R}]{2011ApJ...740..110A}, \object{NGC 6811} \citep[1\,Gyr,][]{2011ApJ...733L...9M,2019ApJ...879...49C}, \object{NGC 6819} \citep[2.5\,Gyr,][]{2015Natur.517..589M}, \object{Ruprecht 147} \citep[2.7\,Gyr,][]{2020AnA...644A..16G,2020ApJ...904..140C}, and \object{M 67} \citep[4\,Gyr,][]{2016ApJ...823...16B,2022ApJ...938..118D,2023AnA...672A.159G}. 

    Taken together, the defined mass dependence of rotation at the age mileposts provided by clusters defines a skeleton of rotational evolution, enabling the ages of non-cluster stars to be read off by comparison with that skeleton, regardless of the details of spindown of stars of differing mass. For additional context, readers may refer to Chap.\,5 in \cite{2021isma.book.....B}. Although such studies have shown that the underlying spindown relation is more complex than originally thought, the fact that all (late-type main sequence) stars populate a single surface in mass-age-rotation period space remains true, a fundamental underpinning for gyrochronology. Recently, \cite{2020AnA...641A..51F} have provided some evidence for the universality of the relation by finding that the rotational distributions of five roughly coeval (125\,Myr) clusters -- \object{NGC 2516}, \object{Pleiades}, \object{M 35} \citep{2009ApJ...695..679M}, \object{M 50} \citep{2009MNRAS.392.1456I}, and \object{Blanco 1} \citep{2014ApJ...782...29C} -- are essentially indistinguishable. Such similarity was also observed at around 2.5\,Gyr for \object{NGC 6819} and \object{Ruprecht 147} by \cite{2020AnA...644A..16G}.

    What is missing is a robust connection to field stars. As a first step to building this connection, we need to establish that the spindown relation observed in open clusters holds for much more diverse field stars as well. While this appears to be likely, it still requires verification. Exploring this with individual field stars returns us to the original problem that field star ages are difficult or even impossible to derive by other methods. But we can investigate a different option -- one that embodies the middle ground between open clusters and individual field stars, namely, wide binaries.

    In certain significant ways, genuine wide binaries constitute the smallest possible open clusters. They are composed of stars born at the same time from the same molecular cloud%
        \footnote{
            It has been suggested that certain wide binaries could be formed via gravitational capture \citep[][and references therein]{2013AN....334...14D}; our results below do not broadly support such a claim for our sample, although it might still be true for the occasional rare system.}.
    As such, they have the same age and metallicity -- the very advantage leveraged by cluster work -- although the precise values may not be known to us. Furthermore, their spindown should have (in a mass-dependent sense) progressed to an equivalent point. Unlike stars in close binaries, the two components of a wide binary are distant enough to be spatially resolved. This distance also means that the components have not interfered in each other's evolution (e.g., by tidal interactions). The rotation periods of cluster stars have already been shown to define certain mass-dependent patterns, with each cluster forming the rib of a skeleton that defines how rotation changes with both mass and cluster age. In this work, we show that wide binary stars follow the same skeleton defined by cluster rotational evolution and  thereby allow an important rotational connection between clusters and field stars to be recognized. We note that though we compare wide binaries with open clusters, wide binaries are generally not present within clusters \citep{2020MNRAS.496.5176D}. In this sense, they are truly field stars.

    This idea has been explored before, although perhaps not as definitively as we are able to do in this paper. \cite{2007ApJ...669.1167B} demonstrated that the components of the wide binaries $\xi$\,Boo\,AB, 61\,Cyg\,AB, and $\alpha$\,Cen\,AB have consistent gyro ages while finding deviations for 36\,Oph\,ABC. \cite{2008ApJ...687.1264M} expanded this list with wide binaries whose rotation periods were estimated from their activity level to compare them with their proposed spindown description. Overall, they found reasonably good agreement between their predictions and the measured periods. However, there were numerous deviations that were enough to cast reasonable doubt on the validity of the fundamental assumptions. Nearly a decade later, \cite{2017ApJ...835...75J} used \emph{Kepler} data for wide binaries to compare their rotational behavior with extant spindown descriptions. His findings echoed those of \cite{2008ApJ...687.1264M}, and the overall trends appeared to be plausible, but a significant number of binaries deviated from the predictions. Janes' work identified a number of "complexities," such as the redder stars rotating faster, which was something unexpected at the time. Updated samples \citep{2018csss.confE..85J,2019AAS...23325936J} did not help overcome these difficulties. Recently, \citet[][]{2022arXiv221001137S} have followed up with another comparison that traced similar steps based on different, more recent spindown descriptions and again arrived at similar, albeit generally more optimistic, conclusions.

    Taking wide binary coevality and the validity of gyrochronology somewhat at face value, \cite{2016MNRAS.455.4212D} briefly touched on the topic based on the \cite{2008ApJ...687.1264M} description, finding their own results to be on the same level of agreement (see their Fig.\,11). Similarly, \cite{2022ApJ...930...36O} devised a method to compare the accuracy of commonly used spindown models regarding their ability to estimate ages for stars in wide binaries. Finally, \cite{2022ApJ...936..109P} explored fully convective M-dwarfs with the help of wide binaries to constrain their spindown evolution.

    The prevailing impression conveyed by prior wide binary work is that while a large fraction of wide binaries confirm expectations, a substantial number do not. The latter create doubt about the generality of spindown and the validity of gyrochronology. However, there is a common denominator in these studies: all of them compared wide binaries to prescriptions of spindown models rather than directly to open cluster observations. (To a certain extent, they had no choice since relevant observations were either sparse or unavailable, especially toward older ages.)

    We argue that such a comparison is flawed in principle, as recent open cluster studies \citep[][]{2020AnA...644A..16G,2020ApJ...904..140C,2022ApJ...938..118D,2023AnA...672A.159G} have shown that the spindown of older cluster stars deviates strongly from the predictions (which were extrapolated from younger ones). And even among younger clusters, the degree to which the models reproduce the observations is questionable. Consequently, comparing a model that is not able to describe the clusters at a given age to wide binaries is not particularly instructive. Therefore, we present a study that mitigates the effects of inadequate spindown models by directly comparing observations alone, that is, open clusters to wide binaries.

    This paper is structured as follows. In Sect.\,\ref{sec_data}, we describe the construction of our wide binary sample, the open cluster data used, and the framework for the subsequent comparison. This part includes the (re)derival or validation of rotation periods, with illustrative examples of identified issues detailed in Appendix\,\ref{sec_issues}. The comparison between wide binaries and clusters is carried out in Sect.\,\ref{sec_analysis}, and the results are further dissected in Sect.\,\ref{sec_discussion}, followed by some conclusions in Sect.\,\ref{sec_conclusion}.

\section{Sample construction and setup} \label{sec_data}

    This section describes the assembly of the relevant data and the manner of preparing them for the subsequent analysis. Our approach was enabled by three developments:
        (1) the availability of the cluster skeleton for FGKM stars out to 4\,Gyr (the age of the M\,67 cluster); 
        (2) the availability of a large sample of wide binaries from \emph{Gaia} astrometry; and
        (3) the availability of space-based photometry from the \emph{Kepler} and \emph{K2} space mission \citep{2010Sci...327..977B,2014PASP..126..398H}, enabling the rotation periods of many of these wide binaries to be determined.
    The base data consists of a sample of wide binaries with vetted rotation periods, a significant number of them were re-derived, and of a set of open cluster stars with known rotation period measurements.

\subsection{Wide binary sample} \label{sec_binary_data}

    We begin with the wide binary catalog of \citet[][EB21 hereafter]{2021MNRAS.506.2269E}. They assembled a list of 1.3 million wide binaries based on Gaia EDR3 \citep{2020yCat.1350....0G} astrometry of stars within 1000\,pc. This sample overlaps to varying degrees with those of \cite{2008ApJ...687.1264M}, \cite{2017ApJ...835...75J,2018csss.confE..85J,2019AAS...23325936J}, and \cite{2022arXiv221001137S}. They followed a cautious approach in their assembly, dismissing doubtful systems from their sample and providing a list with only high probability wide binary pairs. \citetalias{2021MNRAS.506.2269E} also removed stars appearing in multiple binaries and conducted a dedicated search for and elimination of stars in clusters and groups. This ensured that there would be no overlap between our wide binary sample and open cluster stars. (A significant part of the open cluster sample we make use of in our work is based on \emph{Kepler} and \emph{K2} data.) However, this also means that we lost genuine resolved triple systems -- these systems may benefit from a similar, separate future investigation as performed here. Furthermore, the sample has also likely lost a certain number of genuine binaries that suffered chance alignments with more distant tertiaries. Overall, this precaution reduced the sample size, depriving us of a number of genuine systems. However, we ourselves favor a rather exclusive approach to an inclusive one in constructing our sample, for which we require excellent reliability. Therefore, we are fully aligned with the approach of \citetalias{2021MNRAS.506.2269E}.

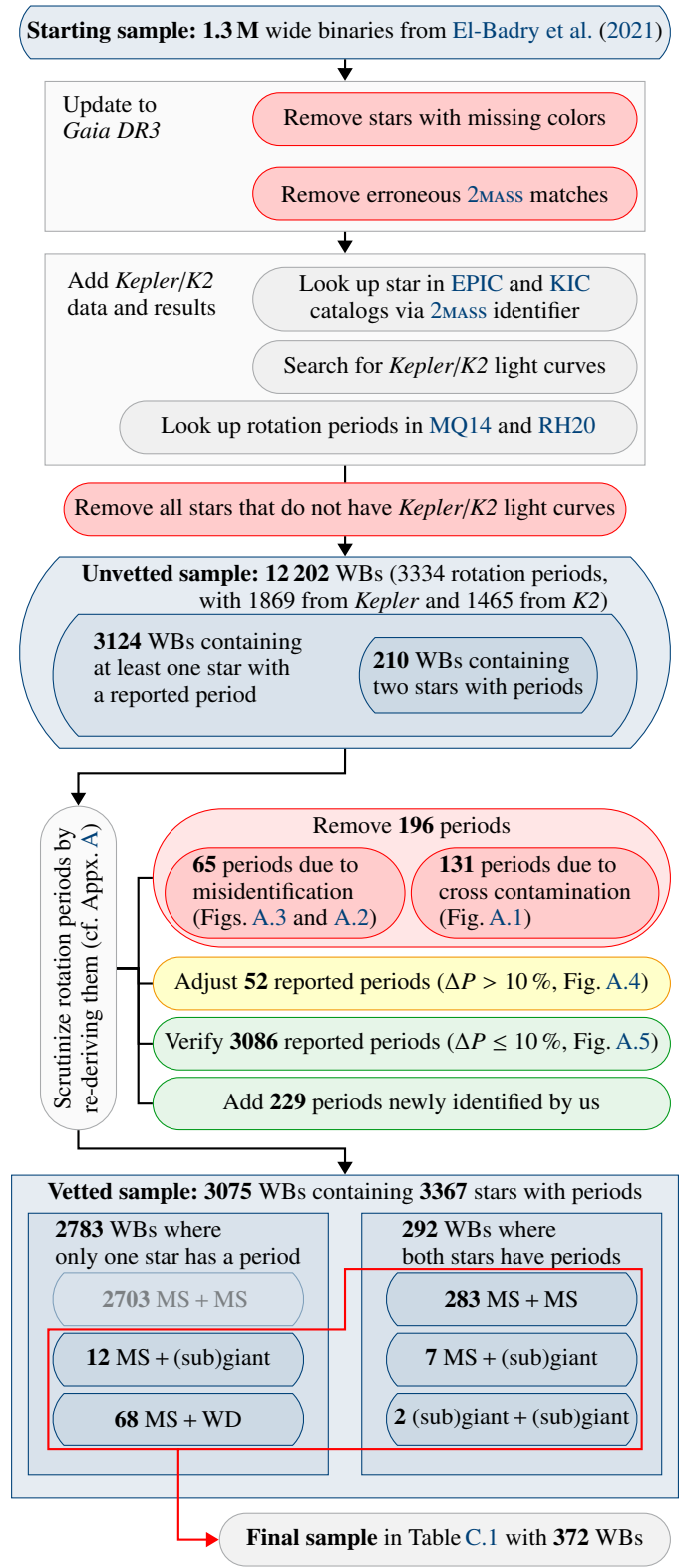
\begin{figure}[ht!]
    \centering
    \begin{minipage}{8.8cm}
        \begin{tikzpicture}\small

    \path(0,-0.15) node[draw=AIPblue,
        rounded rectangle, 
        rounded rectangle arc length=90,
        minimum width=\linewidth,
        fill=AIPblue!10,
        minimum height=.7cm] (eb21) {{\bf Starting sample:} {\bf 1.3\,M} wide binaries from \cite{2021MNRAS.506.2269E}};

    \path(-.0\linewidth,-1.8) node[draw=AIPdarkgrey!50,
        rectangle, 
        minimum width=.9\linewidth,
        fill=AIPlightgrey!20,
        align=left,
        minimum height=2.0cm] (gdr3) {Update to\\ \emph{Gaia DR3} \hspace{6cm} $ $ \\ \\ \\};

    \path(.15\linewidth,-1.3) node[draw=red,
        rounded rectangle,
        fill=red!20,
        align=center,
        minimum width=.6\linewidth,
        minimum height=.7cm] (colors) {Remove stars with missing colors};
    
    \path(.15\linewidth,-2.3) node[draw=red,
        rounded rectangle,
        fill=red!20,
        align=center,
        minimum width=.6\linewidth,
        minimum height=.7cm] (2mass) {Remove erroneous \citetalias{2003yCat.2246....0C} matches};

    \path(-.0\linewidth,-4.5) node[draw=AIPdarkgrey!50,
        rectangle, 
        minimum width=.9\linewidth,
        fill=AIPlightgrey!20,
        align=left,
        minimum height=2.8cm] (kepler) {Add \emph{Kepler}/\emph{K2}\\ data and results \hspace{5.2cm} $ $ \\ \\ \\ \\ \\};

    \path(.15\linewidth,-3.7) node[draw=AIPdarkgrey!50,
        rounded rectangle,
        fill=AIPlightgrey!60,
        align=center,
        minimum width=.6\linewidth,
        minimum height=.7cm] (catalog) {Look up star in \citetalias{2017yCat.4034....0H} and \citetalias{2009yCat.5133....0K}\\ catalogs via \citetalias{2003yCat.2246....0C} identifier};
    
    \path(.15\linewidth,-4.6) node[draw=AIPdarkgrey!50,
        rounded rectangle,
        fill=AIPlightgrey!60,
        align=center,
        minimum width=.6\linewidth,
        minimum height=.7cm] (archive) {Search for \emph{Kepler}/\emph{K2} light curves};

    \path(.05\linewidth,-5.4) node[draw=AIPdarkgrey!50,
        rounded rectangle,
        fill=AIPlightgrey!60,
        align=center,
        minimum width=.8\linewidth,
        minimum height=.7cm] (litperiods) {Look up rotation periods in \citetalias{2014ApJS..211...24M} and \citetalias{2020AnA...635A..43R}};
        
    \path(.0\linewidth,-6.5) node[draw=red,
        rounded rectangle,
        fill=red!20,
        align=center,
        minimum width=.3\linewidth,
        minimum height=.7cm] (lightcurves) {Remove all stars that do not have \emph{Kepler}/\emph{K2} light curves};
    
    \path(0,-8.4) node[draw=AIPblue,
        rounded rectangle, 
        rounded rectangle arc length=90,
        minimum width=\linewidth,
        minimum height=5cm,
        fill=AIPblue!10,
        align=right,
        minimum height=1cm] (sample) {{\bf Unvetted sample:} \hfill {\bf \startsamplebinaries{}} WBs (\alllitperiods{} rotation periods,\\ with \startsampleperiodmq{} from \emph{Kepler} and \startsampleperiodrh{} from \emph{K2}) \\ \alllitperiods{} have a reported period \\ \\ \\ \\};

    \path(-0.00\linewidth,-8.7) node[draw=AIPblue,
        rounded rectangle, 
        rounded rectangle arc length=90,
        minimum width=.9\linewidth,
        minimum height=5cm,
        fill=AIPblue!15,
        align=left,
        minimum height=1cm] (sample2) {\\{\bf\startingbinariesall{}} WBs containing \\at least one star with \\ a reported period \hspace{.5\linewidth}$ $\\};
    
    \path(.2\linewidth,-8.7) node[draw=AIPblue,
        rounded rectangle, 
        rounded rectangle arc length=90,
        minimum width=.3\linewidth,
        fill=AIPblue!20,
        align=left,
        minimum height=1cm] (sample3) {{\bf\binariestwoperiods{}} WBs containing\\ two stars with periods};

    \path(-.4\linewidth,-12.6) node[draw=AIPdarkgrey!50,
        rounded rectangle, 
        minimum width=.1\linewidth,
        fill=AIPlightgrey!20,
        align=center,rotate=90,
        minimum height=1cm] (sample4) {Scrutinize rotation periods by\\re-deriving them (cf. Appx.\,\ref{sec_issues})};

    \path(.1\linewidth,-11.4) node[draw=red,
        rounded rectangle, 
        minimum width=.8\linewidth,
        fill=red!10,
        align=left,
        minimum height=1cm] (remove) {Remove {\bf \alldismissed{}} periods\\ \\ \\ \\ };
    
    \path(-.09\linewidth,-11.6) node[draw=red,
        rounded rectangle, 
        minimum width=.3\linewidth,
        fill=red!20,
        align=left,
        minimum height=1cm] (miss) {{\bf \allmissingtoall{}} periods due to\\ misidentification\\ (Figs.\,\ref{fig_lightcurve_trending} and \ref{fig_mix_periods})};

    \path(.285\linewidth,-11.6) node[draw=red,
        rounded rectangle, 
        minimum width=.3\linewidth,
        fill=red!20,
        align=left,
        minimum height=1cm] (cross) {{\bf \brokenbinaries{}} periods due to\\ cross contamination\\ (Fig.\,\ref{fig_lightcurve_contamination})};

    \path(.1\linewidth,-12.8) node[draw=AIPorange,
        rounded rectangle, 
        minimum width=.8\linewidth,
        fill=tikzyellow!30,
        align=left,
        minimum height=.7cm] (change) {Adjust {\bf \changedperiods{}} reported periods ($\Delta P>10\,\%$, Fig.\,\ref{fig_adjust_period1})};

    \path(.1\linewidth,-13.6) node[draw=tikzgreen,
        rounded rectangle, 
        minimum width=.8\linewidth,
        fill=tikzgreen!10,
        align=left,
        minimum height=.7cm] (verify) {Verify {\bf \closeperiods{}} reported periods ($\Delta P\leq10\,\%$, Fig.\,\ref{fig_adjust_period2})};

    \path(.1\linewidth,-14.4) node[draw=tikzgreen,
        rounded rectangle, 
        minimum width=.8\linewidth,
        fill=tikzgreen!10,
        align=left,
        minimum height=.7cm] (add) {Add {\bf \alladditonaltoall{}} periods newly identified by us};

    \path(0,-17.5) node[draw=AIPblue,
        rectangle, 
        rounded rectangle arc length=90,
        minimum width=\linewidth,
        fill=AIPblue!10,
        align=left,
        minimum height=.7cm] (final) {{\bf Vetted sample:} {\bf\startsamplebinariestotal{}} WBs containing {\bf\startsamplebinarieswperiods{}} stars with periods \\ \\ \\ \\ \\ \\ \\ \\ \\ \\ \\};

    \path(.25\linewidth,-17.6) node[draw=AIPblue,
        rectangle, 
        minimum width=.45\linewidth,
        fill=AIPblue!15,
        align=left,
        minimum height=.7cm] (double) {{\bf \alltwoperiodwidebinaries{}} WBs where\\ both stars have periods \\ \vspace{2cm} \\ $ $};

    \path(-.25\linewidth,-17.6) node[draw=AIPblue,
        rectangle, 
        minimum width=.45\linewidth,
        fill=AIPblue!15,
        align=left,
        minimum height=.7cm] (single) {{\bf \startsamplebinarieswonep{}} WBs where\\ only one star has a period \\ \vspace{2cm} \\ $ $};

    \path(-.25\linewidth,-17.0) node[draw=AIPblue!50,
        rounded rectangle, 
        rounded rectangle arc length=90,
        minimum width=.4\linewidth,
        fill=AIPblue!15,
        align=left,
        minimum height=.7cm] (single) {\color{AIPdarkgrey!75}{\bf \msmsoneperiod{}} MS\,+\,MS };

    \path(-.25\linewidth,-17.8) node[draw=AIPblue,
        rounded rectangle, 
        rounded rectangle arc length=90,
        minimum width=.4\linewidth,
        fill=AIPblue!20,
        align=left,
        minimum height=.7cm] (single) {{\bf \evolvedwithoneperiod{}} MS\,+\,(sub)giant};

    \path(-.25\linewidth,-18.6) node[draw=AIPblue,
        rounded rectangle, 
        rounded rectangle arc length=90,
        minimum width=.4\linewidth,
        fill=AIPblue!20,
        align=left,
        minimum height=.7cm] (single) {{\bf \binarywd{}} MS\,+\,WD};

    \path(.25\linewidth,-17.0) node[draw=AIPblue,
        rounded rectangle, 
        rounded rectangle arc length=90,
        minimum width=.4\linewidth,
        fill=AIPblue!20,
        align=left,
        minimum height=.7cm] (single) {{\bf \mswidebinaries{}} MS\,+\,MS};

    \path(.25\linewidth,-17.8) node[draw=AIPblue,
        rounded rectangle, 
        rounded rectangle arc length=90,
        minimum width=.4\linewidth,
        fill=AIPblue!20,
        align=left,
        minimum height=.7cm] (single) {{\bf \evolvedwithtwoperiods{}} MS\,+\,(sub)giant};

    \path(.25\linewidth,-18.6) node[draw=AIPblue,
        rounded rectangle, 
        rounded rectangle arc length=90,
        minimum width=.4\linewidth,
        fill=AIPblue!20,
        align=left,
        minimum height=.7cm] (single) {{\bf 2} (sub)giant\,+\,(sub)giant};

    \path(.15\linewidth,-20.2) node[draw=AIPdarkgrey!50,
        rounded rectangle, 
        minimum width=.7\linewidth,
        fill=AIPlightgrey!50,
        align=left,
        minimum height=.7cm] (table) {{\bf Final sample} in Table\,\ref{tab_period_sample} with {\bf \allbinaries{}} WBs};

    \draw[-Triangle,thick,draw=red] (-.25\linewidth,-19.0) -- (.445\linewidth,-19.0) |- (-.0\linewidth,-16.6) |- (-.445\linewidth,-17.4) |- (-.25\linewidth,-19.0) -- (-.25\linewidth,-20.2) -- (table);    
    
    \draw[-Triangle,thick] (eb21) -- (gdr3);
    \draw[-Triangle,thick] (gdr3) -- (kepler);
    \draw[-,thick] (kepler) -- (lightcurves);
    \draw[-Triangle,thick] (lightcurves) -- (sample);
    \draw[-Triangle,thick] (sample) -- +(0,-1.6) -| (sample4);
    \draw[-Triangle,thick] (sample4) -- +(0,-2.4) -| (final);
    
    \draw[-,thick] (add)    -| (-.31\linewidth,-12.6) -- (sample4);
    \draw[-,thick] (remove) -| (-.31\linewidth,-12.6) -- (sample4);
    \draw[-,thick] (verify) -| (-.31\linewidth,-12.6) -- (sample4);
    \draw[-,thick] (change) -| (-.31\linewidth,-12.6) -- (sample4);
 
\end{tikzpicture}\normalfont
    \end{minipage}
    \caption{
        Flowchart visualizing the steps of the wide binary sample selection process detailed in Sect.\,\ref{sec_binary_data}. (WBs\,=\,wide binaries)
    }
    \label{fig_flowchart}
\end{figure}

    The step-by-step process of assembling our sample is illustrated in Fig.\,\ref{fig_flowchart} and detailed below. We updated the stellar parameters of the \citetalias{2021MNRAS.506.2269E} sample with Gaia DR3 \cite[GDR3,][]{2022arXiv220800211G} measurements. Because the faintest stars often do not have $G_\mathrm{BP}$ and $G_\mathrm{RP}$ magnitudes, we dismissed those stars from the sample. Furthermore, based on $G_\mathrm{BP}-G_\mathrm{RP}$ and $G-G_\mathrm{RP}$, we removed stars whose color-magnitude diagram (CMD) positions are inconsistent to avoid stars with problematic photometry. \citetalias{2022arXiv220800211G} includes a match to The Two Micron All Sky Survey \citep[\textsc{2mass};][]{2006AJ....131.1163S} catalog, which we adopted as is. We note that the \citetalias{2022arXiv220800211G} match to \citetalias{2003yCat.2246....0C} is not perfect. There are numerous binaries where both component \citetalias{2022arXiv220800211G} sources have been matched to the same \citetalias{2003yCat.2246....0C} source; again, we dismissed such binaries from our sample.

    Based on the \citetalias{2003yCat.2246....0C} identifiers, we looked up the stars in the Kepler Input Catalog \citep[KIC;][]{2009yCat.5133....0K} and the K2 Ecliptic Plane Input Catalog \citep[EPIC;][]{2017yCat.4034....0H}. In turn, those catalog counterparts were used to match stars to their \emph{Kepler} and \emph{K2} archive light curves and to find them in the rotation period samples of \citet[][$\sim$34\,000 rotation periods from \emph{Kepler}; MQ14 hereafter]{2014ApJS..211...24M} and \citet[][$\sim$30\,000 rotation periods from \emph{K2}; RH20 hereafter]{2020AnA...635A..43R}. We note that while \citetalias{2020AnA...635A..43R} report individual periods for each \emph{K2} campaign a particular star was observed in, those periods can be inconsistent%
        \footnote{
            For example, \object{EPIC 211638150}, for which \citetalias{2020AnA...635A..43R} reports $P_\mathrm{C05}=26.8\pm4.8$\,d, $P_\mathrm{C16}=20.6\pm2.2$\,d, and $P_\mathrm{C18}=22.7\pm2.9$\,d.}.
    In case of multiple reported periods for a given star, we adopted the one with the highest normalized peak\footnote{See ``Hpeak'' column in their catalog at \url{J/A+A/635/A43/table2}.}. Here, we only retained wide binaries for which at least one star has an associated light curve. This procedure provided us with a sample of \startsamplebinaries{} wide binaries (\startsamplestars{} stars, with \startsampleperiodmqall{} from \emph{Kepler} and \startsampleperiodrhall{} from \emph{K2}). Only a minority of these stars (\alllitperiods{}, with \startsampleperiodmq{} from \emph{Kepler} and \startsampleperiodrh{} from \emph{K2}) have a reported period. Those stars belong to \startingbinariesall{} wide binaries, with \binariestwoperiods{} of them being systems for which both stars have an identified rotation period.

    One of the criteria we demanded for this work was the validity of the rotation periods used. This encouraged us to inspect the periods in some detail. It soon became obvious that a certain fraction of the rotation periods reported in \citetalias{2014ApJS..211...24M} and \citetalias{2020AnA...635A..43R} are not as reliable as we had hoped. We identified problems that render a reported rotation period unusable. These issues appear to originate in one of two sources:
        (1) from light curve creation -- here there is cross contamination from nearby stars (removes \brokenbinaries{} periods); and 
        (2) from period detection -- here there is misidentification of non-periodic variability and/or remaining data systematics as a periodic signal (removes \allmissingtoall{} periods).
    We illustrate these problems in more detail in Appendix Sect.\,\ref{sec_issues}. These issues led us to manually inspect all stars, along with their on-sky surroundings, their periods, and light curves, in our sample to either verify ($\Delta P_\mathrm{rot}\leq10\,\%$, \closeperiods{} stars), adjust ($\Delta P_\mathrm{rot}>10\,\%$, \changedperiods{} stars), or altogether dismiss (\alldismissed{} stars) the previously reported rotation periods. This inspection involved a comparison with light curves from similar\footnote{That is, comparable in color and brightness.} stars in the surroundings to identify common trends. The \emph{K2} targets were found to be particularly affected by this issue. For this process, we downloaded the \emph{Kepler} and \emph{K2} light curves for all relevant stars. Where available, we used the K2 Systematics Correction\footnote{\url{archive.stsci.edu/prepds/k2sc/}} \citep[\texttt{k2sc},][]{2016MNRAS.459.2408A} and EPIC Variability Extraction and Removal for Exoplanet Science Targets\footnote{\url{archive.stsci.edu/hlsp/everest}} \citep[\citetalias{2016AJ....152..100L},][]{2016AJ....152..100L,2018AJ....156...99L} light curves for additional validation. We also inspected the immediate surroundings of a star for crowding relative  to the comparatively low spatial resolution of \emph{Kepler}. When the origin of the observable variability was in doubt, we rejected the star from our sample.

    We were also able to add rotation periods that had not been identified in \citetalias{2014ApJS..211...24M} or \citetalias{2020AnA...635A..43R} for an additional \alladditonaltoall{} stars (\alladditonaltomq{} from \emph{Kepler} and \alladditonaltorh{} from \emph{K2}). These were derived using procedures described in detail in \citet[][\citetalias{2023AnA...672A.159G} hereafter]{2023AnA...672A.159G}. We note that we suppressed the period errors in all figures for visibility reasons. However, they are listed in Table\,\ref{tab_period_sample}. Typically, the error is $P_\mathrm{err} \sim 0.1 P_\mathrm{rot}$.

\begin{figure}[ht!]
    \centering
    \includegraphics[width=8.8cm]{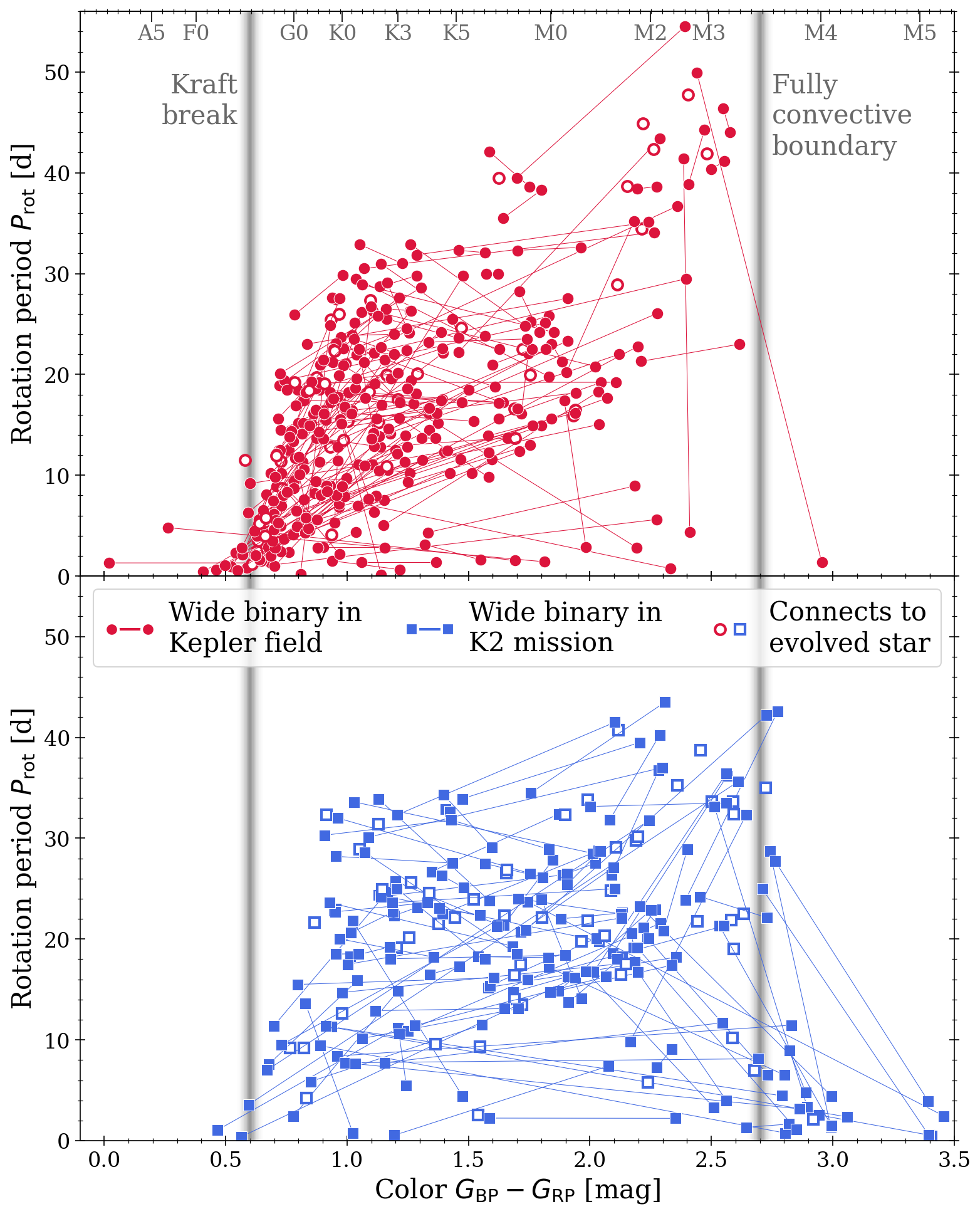}
    \caption{
        Our wide binary sample split into the Kepler field and K2 subsamples. The two components of each binary are connected by straight lines. The Kraft break (beyond which more massive stars lack an outer convective envelope and thus do not experience significant magnetic braking) and the boundary between partially convective and fully convective stars are marked. Unfilled symbols indicate the main sequence components of wide binaries that contain evolved stars, which typically do not have measured periods.
    }
    \label{fig_sample_overview_cmdcpd}
\end{figure}

    We thus arrived at a sample of \startsamplebinarieswperiods{} stars with rotation periods that belong to stars in \startsamplebinariestotal{} wide binaries. Of our sample of wide binaries, \mswidebinaries{}  are composed of two main sequence stars where both have rotation periods. Figure\,\ref{fig_sample_overview_cmdcpd} illustrates their distribution in the color-period diagram (CPD), and Table\,\ref{tab_period_sample} lists an excerpt of our full sample. The figure helps to visualize the different emphases of the \emph{Kepler} and \emph{K2} missions: \emph{Kepler} favored G-type stars, whereas \emph{K2} emphasized redder spectral types. However, in all subsequent matters, we do not distinguish between \emph{Kepler} and \emph{K2} targets, and we treat them as equivalent.

    There are of course a large number (\startsamplebinarieswonep{}) of additional wide binaries for which only one of the component stars has a measured rotation period. The majority of these are composed of two main sequence (MS) stars. The periods for these stars lie in the same regions of the CPD as our sample, but the absence of a companion period prevented us from making an effective comparison between the two\footnote{While we used evolved stars where an isochrone can be used to constrain the system age reasonably, the uncertainty on a main sequence isochrone age is too large to be useful. In fact, this is the fundamental motivation for gyrochronology itself.}. Accordingly, we did not consider them further in this work. However, there is a certain type of wide binary system with only one period that retained interest for us, namely, those where the star without a period is somewhat evolved. In such cases, the stars have left the main sequence, and their positions in a CMD correlate strongly with their ages. This means that the evolved component in these systems can help us verify the (rotation period-based) age of the non-evolved MS component. We address the \allevolved{} systems that fall under this case separately below. For seven{} systems, the evolved stars have rotation periods in addition to the main sequence components, but we lumped them together with the other systems that contain an evolved component here. The process described above left us with \allbinaries{} wide binaries for further investigation.

    We note that we did not have a consistent source of reddening estimates for the wide binary stars, as the values reported in \citetalias{2022arXiv220800211G} were also rather unreliable. Therefore, we were left to use the reddened colors and carried out the comparison with this caveat in mind. \cite{2020ApJ...904..140C} have shown that the median reddening for the \emph{Kepler} field stars is $E(B-V)=0.04$\,mag/kpc. Given that our sample is fully within 1\,kpc, it is unlikely that reddening is a significant issue here.

\subsection{Open cluster data} \label{sec_cluster_data}

    We next discuss how we constructed a comparison sample based on the open cluster work currently available. The rationale for this approach was our desire to be guided by the data alone, where possible, instead of performing a comparison between the wide binaries and models of rotational evolution, as has been typical in prior work. Our task was enabled by recent work, including our own, in older open clusters that was unavailable before. In particular, we investigated the following question: If one component of a wide binary has a rotation period that places it on the rotational sequence of a cluster, is it likely that the other component is also on the cluster sequence?
    Curiously, this appears to be extremely likely, and this motivated us to create a set of groups that roughly trace out the cluster sequences and allow the binaries to each be associated with a particular cluster, a group of clusters, or the inter-cluster regions between them\footnote{This is parallel to the usage of the `group' terminology in geological stratigraphy.}.

\setlength{\tabcolsep}{5pt}
\begin{table}[h!]
    \caption{Overview of the open cluster sample.}
    \label{tab_cluster_compilation}
    \centering
    {\small
    \begin{tabular}{llccccc}
    \hline\hline
    \\[-0.8em]
    Age     & Cluster      & Age  & [Fe/H]  & $A_V$  & Sample            & $N_\text{star}$ \\ 
    Group   &              &[Gyr] &       & [mag]  & ref.              &    \\ 
    \hline 
    \\[-0.8em]
    1       & \object{Pleiades}     & 0.15 &  0.032  & 0.168 & 1 &   759 \\ 
    \\[-0.7em]
            & \object{Blanco 1}     & 0.15 & -0.016  & 0.064 & 2  &   127 \\ 
    \\[-0.7em]
            & \object{NGC 2516}     & 0.15 & -0.008  & 0.449 & 3  &   308 \\ 
    \\[-0.8em]
            & \object{M 35}         & 0.15 & -0.123  & 0.883 & 4 &   441 \\             
    \\[-0.7em]
            & \object{NGC 3532}     & 0.3  &  0.050  & 0.136 & 5  &  279 \\ 
    \hline                    
    \\[-0.8em]
    2       & \object{Hyades}       & 0.65 &  0.149  & 0.034 &  6 &   23 \\ 
            &                       &      &         &  &  7 &  237 \\ 
    \\[-0.7em]
            & \object{Praesepe}     & 0.7  &  0.196  & 0.032 & 7  &  743 \\  
    \\[-0.7em]
            & \object{NGC 6811}     &  1.0 &  0.032  & 0.213 & 8  &   71 \\ 
            &                       &      &         &  & 9  &  171 \\ 
    \hline                    
    \\[-0.8em]
    3       & \object{NGC 752}      &  1.4 & -0.037  & 0.159 & 10  &   12 \\ 
    \hline                    
    \\[-0.8em]
    4       & \object{NGC 6819}     &  2.5 &  0.093  & 0.487 & 11  &   30 \\ 
    \\[-0.7em]
            & \object{Ruprecht 147} &  2.7 &  0.089  & 0.292 & 12  &   32 \\  
            &                       &      &         &  & 13  &   35 \\ 
    \hline  
    \\[-0.8em]
    5       & \multicolumn{5}{l}{\it No open clusters for $2.7<t<4$\,Gyr} \\ 
    \hline                    
    \\[-0.8em]
    6       & \object{M 67}         &   4.0 &  0.072  & 0.127 &  14  &  20 \\ 
            &                       &       &         &  &  15  &  64 \\ 
            &                       &       &         &  &  16  &  47 \\ 
    \hline
\end{tabular}
    }
    \tablefoot{
        Age groups refer to the rough age classification from Sect.\,\ref{sec_analysis}. The [Fe/H] and $A_V$ for all clusters was taken from \cite{2021MNRAS.504..356D}.}
    \tablebib{
        (1)~\cite{2016AJ....152..113R};
        (2)~\cite{2020MNRAS.492.1008G};
        (3)~\cite{2020AnA...641A..51F}:
        (4)~\cite{2009ApJ...695..679M};
        (5)~\cite{2021AnA...652A..60F};
        (6)~\cite{1987ApJ...321..459R};
        (7)~\cite{2019ApJ...879..100D};
        (8)~\cite{2011ApJ...733L...9M};
        (9)~\cite{2019ApJ...879...49C};
        (10)~\cite{2018ApJ...862...33A};
        (11)~\cite{2015Natur.517..589M};
        (12)~\cite{2020AnA...644A..16G};
        (13)~\cite{2020ApJ...904..140C};
        (14)~\cite{2016ApJ...823...16B};
        (15)~\cite{2022ApJ...938..118D};
        (16)~\cite{2023AnA...672A.159G}
    }
\end{table}
\setlength{\tabcolsep}{6pt}

\begin{figure}[ht!]
    \centering
    \includegraphics[width=8.8cm]{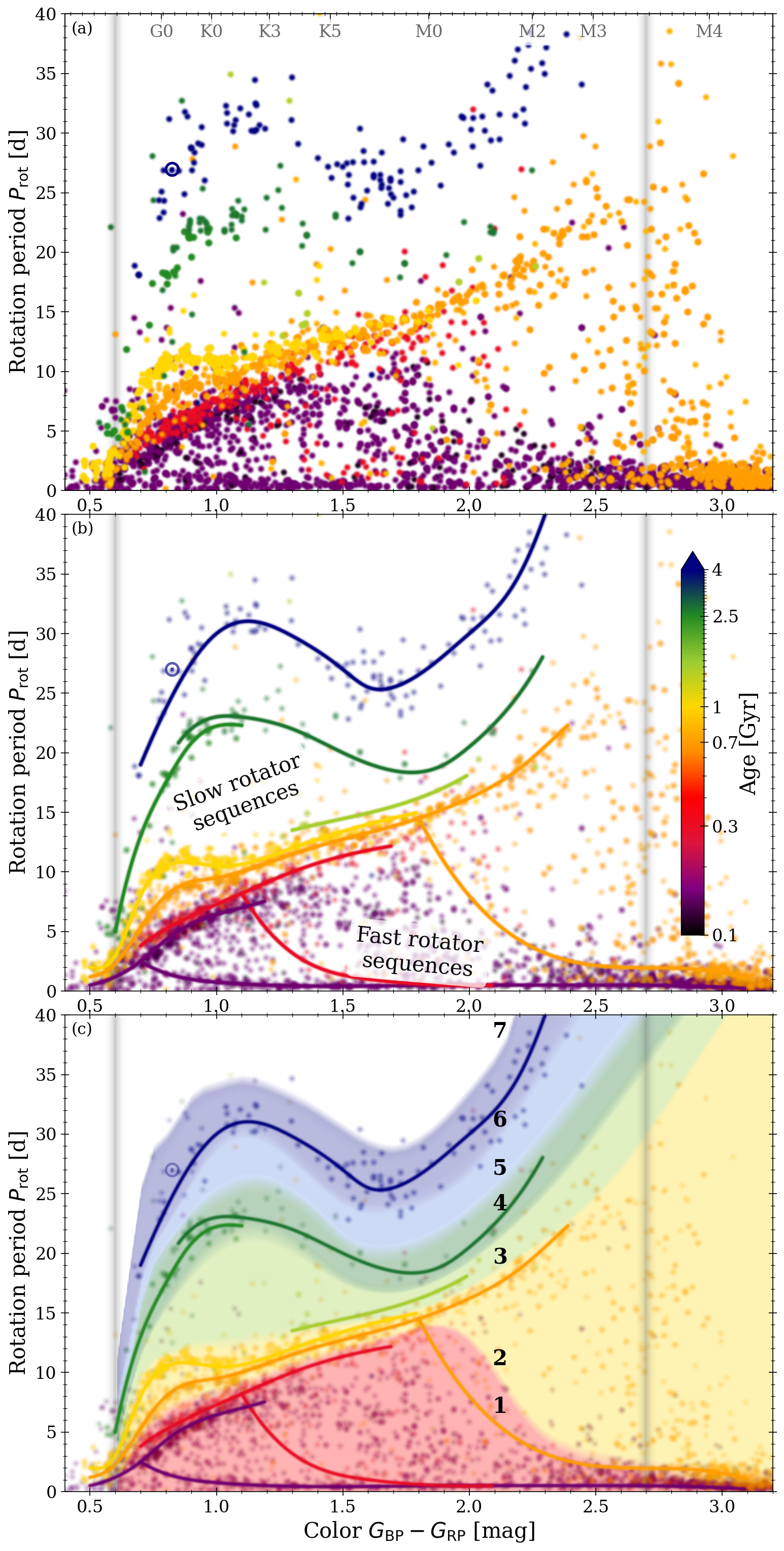}
    \caption{
        Color-period diagrams for our sample of open cluster stars. Panel (a): All the stars from our open cluster compilation (cf. Table\,\ref{tab_cluster_compilation}). Panel (b): Same stars but overplotted with a manual indication of the sequences followed by stars of the same age (the cluster sequences), with slow and fast rotator sequences where available. Panel (c): How cluster sequences and inter-cluster gaps were used to divide the continuous space of the CPD into groups 1--7 (cf. Sect.\,\ref{sec_age_binning}). Stars, sequences, and groups are all color coded by age (see color bar in panel (b), based on the cluster ages in Table\,\ref{tab_cluster_compilation}). The Sun is shown with its usual symbol. As in Fig.\,\ref{fig_sample_overview_cmdcpd}, vertical gray lines indicate the Kraft break and the fully convective boundary.
    }
    \label{fig_cluster_cpd}
\end{figure}

    We selected a number of open clusters ranging in age from the zero age main sequence (ZAMS) to 4\,Gyr as our comparison sample. They are listed in Table\,\ref{tab_cluster_compilation}. This sample was intended to cover the age and color range rather than to be complete with respect to the specific open clusters covered or to studies of the listed clusters. Nonetheless, it is sufficient for illustrating the cluster sequences at a given age to the extent of current availability in the literature. Overlap between different studies on the same open cluster are permitted, and we did not remove stars that occur more than once in the sample. From each source catalog, we only adopted the rotation period and the identifiers used for the stars and matched all to \citetalias{2022arXiv220800211G}. This resulted in a consistent set of photometry for all wide binaries and open clusters. The accumulated data for 12 open clusters is shown in panel (a) of Fig.\,\ref{fig_cluster_cpd}.
    We obtained the $E(B-V)$ reddening for each cluster from the compilation of \cite{2021MNRAS.504..356D} and calculated the reddening in Gaia colors via the prescription of \cite{2018MNRAS.479L.102C}:
    \begin{equation}\label{eq_reddening}
        E(G_\mathrm{BP}-G_\mathrm{RP}) = 1.337 \cdot E(B-V).
    \end{equation} 
    In all matters, we always used dereddened colors for the cluster stars.

    We note that we did not subject the stars from the open cluster sample to the same scrutiny as the wide binary stars. There are outliers in the individual cluster samples, but the very fact that we can identify them as outliers is already telling, as the clusters are defined well enough that outliers are actually visible. It is those sequences themselves that we ultimately compared our wide binaries to.

    In panel (b) of Fig.\,\ref{fig_cluster_cpd}, we have drawn fiducial lines by eye through the cluster sequences. Keeping in mind that there is a continuous color-period space (or color-age space), we have distinguished the cluster age groups with different colors in this figure (cf. panel c) and in the following ones. Groups 1, 2, 4, and 6 are associated directly with the open cluster sequences (see Table\,\ref{tab_cluster_compilation} and Sect.\,\ref{sec_age_binning} for details). Because of the large age gap between the clusters, we have inserted additional groups. Group\,3 was placed in the gap between Praesepe/NGC\,6811 (2) and NGC\,6819/Ruprecht\,147 (4), whereas group\,5 was placed between NGC\,6819/Ruprecht\,147 (4) and M\,67 (6). Finally, we added a group (7) for wide binary systems older than M\,67. The result is effectively a stratification of what is actually a continuous distribution of stars in the color-period space (cf. panel c of Fig.\,\ref{fig_cluster_cpd}). However, as with certain other groupings of continuous distributions into discrete bins,\footnote{The colors assigned to different segments of the rainbow and demographic cohorts (generations) are obvious examples.} there seems to be a distinct benefit to the grouping, as we show below.

\subsection{Grouping of the open cluster distribution}\label{sec_age_binning}

    We defined the set of seven age groups in the CPD whose mass-dependent range is dictated by the open clusters, including the sequences themselves and the intervening gaps. The definitions of the groups are detailed in the following paragraphs.
    
    Group\,1 contains main sequence clusters younger than 500\,Myr. These clusters contain significant numbers of stars occupying the fast rotator sequence. The group incorporates the ZAMS clusters Pleiades, Blanco\,1, NGC\,2516, M\,35, and the roughly 300\,Myr-old cluster NGC\,3532.
    
    Group\,2 includes clusters with ages ranging from $0.5$\,Gyr to $1$\,Gyr, thereby containing Hyades, Praesepe, and NGC\,6811. The fast rotators in these clusters (with the exception of M-dwarfs) have converged to the slow rotator sequence.
    
    Group\,3 ranges from (above) 1 to (below) 2.5\,Gyr and marks the first inter-cluster gap in the cluster distribution. Only NGC\,752 is available for this group, but it barely contains any stars with known periods to define a sequence%
        \footnote{\cite{2020ApJ...904..140C} have noted that of the 12 rotators reported in \cite{2018ApJ...862...33A}, four are non-members post Gaia\,DR2 and many of the rest are binaries.}. 

    Group\,4 covers the 2.5\,Gyr to 2.8\,Gyr region and contains the more evolved clusters NGC\,6819 and Ruprecht\,147.

    Group\,5 spans the age range from (above) 2.8\,Gyr to (below) 4\,Gyr and marks the second inter-cluster gap. No open cluster study in this age range is currently available.

    Group\,6, around an age of approximately 4\,Gyr, is formed solely by the oldest open cluster studied with respect to rotation: M\,67. 
    
    Group\,7 includes all wide binaries likely older than 4\,Gyr where no open cluster has been explored to date.
    
    Panel (c) of Fig.\,\ref{fig_cluster_cpd} shows the (approximate) ranges covered by these groups in comparison with the open cluster data. We acknowledge some leeway in the group stratification. For instance, age group\,6 was set around 4\,Gyr but also includes the Sun \citep[4.56\,Gyr,][cf. also the CPD position of the Sun in comparison with the width of M\,67 sequence]{2002Sci...297.1678A}. There is also some overlap in the CPD among the fast-rotating stars.

\section{Grouping the wide binaries} \label{sec_analysis}

    In this section, we compare the distribution of rotation periods as seen in the wide binaries with the age groups we have defined based on the open clusters. We wished to investigate the consistency between the wide binary components with respect to the groups and explore the agreements and deviations. There is one caveat in this comparison; while the colors of the clusters are dereddened, the colors of the binary stars are not, and they are subject to an unknown amount of reddening. This effect is likely small for most stars, but it could affect the grouping of the more distant or the bluest stars in our sample (due to the strong color dependence of the cluster rotational sequences for $G_\text{BP}-G_\text{RP}<0.8$\,mag). The comparison below was carried out with this in mind, occasionally permitting equal reddening for both binary components to be considered consistent.

    First, we assigned each wide binary to one of our age groups. For this procedure, we designated the stars of the individual wide binaries as the leading (\primary{}) and trailing (\secondary{}) components. The \primary{} component was placed in the CPD and assigned to a group based on its $G_{BP}-G_{RP}$ color and period $P_\mathrm{rot}$. With that, we essentially used gyrochronology and assigned it a gyro age, an age based on its rotation period and color, albeit a rather rough one. The choice of which star in a particular wide binary was designated as \primary{} was based on which star had a CPD position that allowed for a better estimation of the age (group). Furthermore, we avoided stars in a somewhat evolved state and, when possible, very fast rotating stars. Typically, this meant that we adopted the star whose color is in the range $0.8<G_\text{BP}-G_\text{RP}<1.8$ since the age groups allow for the best distinction in that region. Crucially, we avoided late-F and early G-stars as leading components because of the strong color dependence in this region and the (likely small) uncertainties arising from the unknown reddening.

\begin{figure}[ht!]
    \centering
    \includegraphics[width=8.8cm]{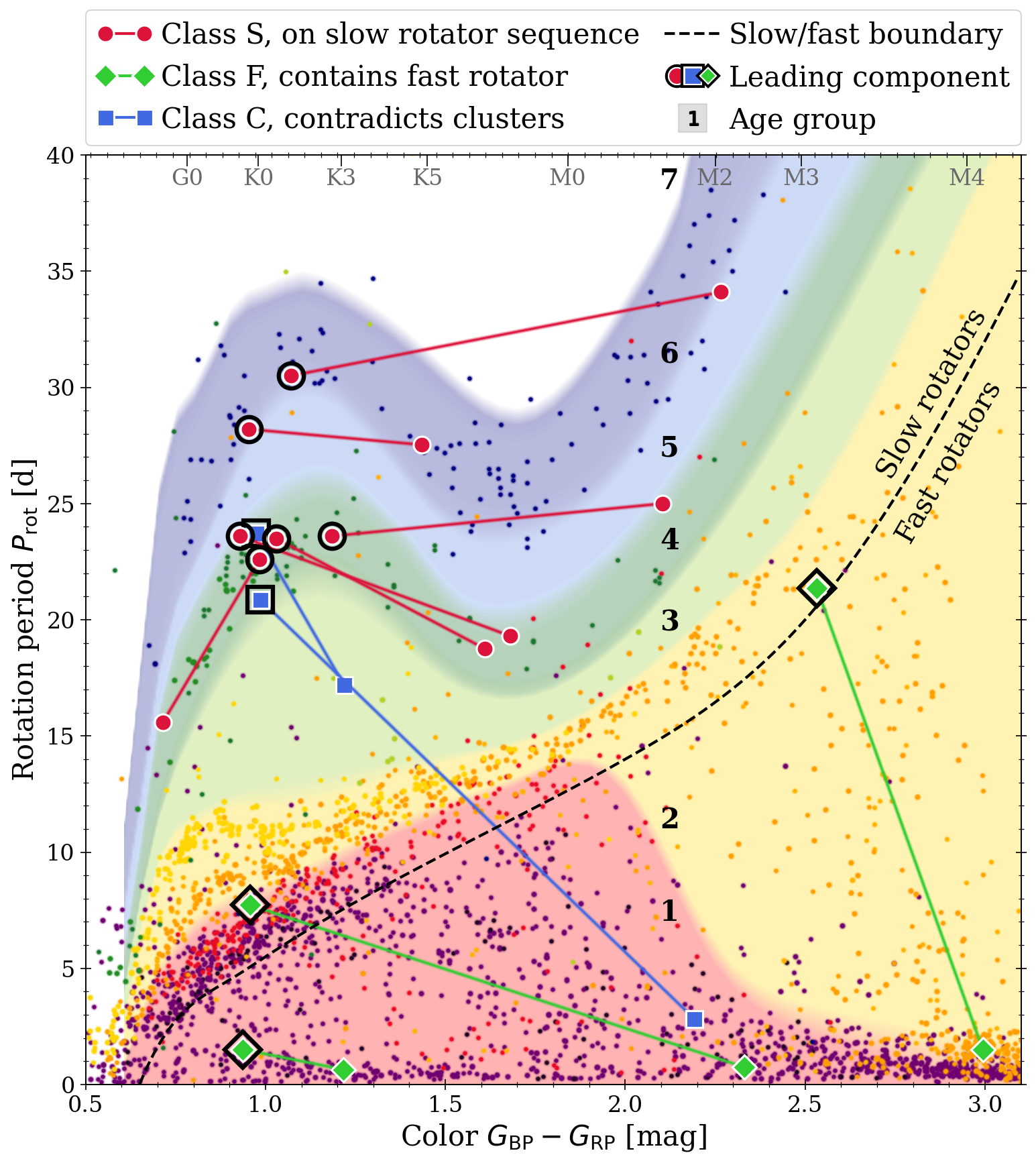}
    \caption{
        Color-period diagram defining age groups (colored regions) and illustrating the classification of our wide binaries. Color coding roughly follows Fig.\,\ref{fig_cluster_cpd} and corresponds to the cluster stars plotted in the background (small dots). Age group numbers (1\,--\,7) correspond to Table\,\ref{tab_cluster_compilation}. Eleven sample wide binaries are overplotted. Red circles indicate binaries agreeing with the slow rotator sequences, green squares are those containing a fast rotator, and blue boxes mark those altogether disagreeing with the cluster behavior. Stars of each wide binary that were preferentially used to assign them to an age bin (i.e., the \primary{} components) are outlined in black. The dashed line denotes the approximate boundary used to distinguish between fast and slow rotators (cf. the sequences indicated in panel (b) of Fig.\,\ref{fig_cluster_cpd}). The systems plotted are listed in Table\,\ref{tab_period_sample}.
    }
    \label{fig_age_bins}
\end{figure}

    Figure\,\ref{fig_age_bins} illustrates this process on a few selected wide binary pairs. Encircled stars mark the \primary{} component for the selection process. We note that wide binary systems were selected for the figure to allow good visualization. This means we selected \primary{} components of similar colors that span several age groups and connect to different areas of the CPD. This selection essentially covers the diversity of the systems we found.

    When all \primary{} components were set, we could then take the \secondary{} components, place them in the CPD, and see how their locations compared with the age groups assigned to the \primary{} components. Depending on the result of this comparison, we classified the wide binary pair itself as either \classS, \classF, or \classC, depending on the content\footnote{Naming derived from slow rotator (\classS){}, fast rotator (\classF{}), contradicting (\classC{}), evolved (\classE{}), and white dwarf (\classW{}).\label{fn_classes}}. For the classification of binary systems composed of two MS stars, each with a listed rotation period, we used the criteria described in the following paragraphs.
    
    If \secondary{} falls reasonably well on the slow rotator sequence in the same age group as \primary{}, we marked the whole system as \classS. Such systems are shown with connected red circles in {Fig.\,\ref{fig_age_bins}.}
    
    If \secondary{} is located where the open clusters matching \primary{}'s age group indicate a fast rotator population, then we classified the system as \classF. All of these binaries are from age groups\,1 and 2 because only such clusters have a population of fast rotators. Such systems are shown with (connected) green squares in the figure.
    
    All other systems, meaning those whose \secondary{} components are main sequence stars and do not fall in a region that is populated by cluster stars of \primary{}'s age group, were classified as \classC{}. They are indicated with connected blue boxes in {Fig.\,\ref{fig_age_bins}.
    
    Age group\,7 is special in that there are no cluster predictions and it contains all systems that are arguably older than age group\,6 (4\,Gyr). Since we could not fully classify systems in this group, all systems in group 7 were designated as \classS. However, we note that we later investigate these systems separately in Sect.\,\ref{sec_extrapolate}. Exceptions to this classification were obvious contradicting systems that contain one star in age group\,7 and one in another group. Such instances were accordingly classified as \classC.
    
    Several systems in age groups\,1 and 2 are composed of two fast rotators and were thus classified as \classF{}.
    
    We note that it is more difficult to identify rotationally deviant component stars in wide binaries as compared with clusters where the sheer numbers of other stars on the slow rotator sequences make outliers stand out.

    Systems whose \secondary{} component is not a main sequence star but has evolved past it were classified based on \secondary{}'s state. If \secondary{} is a (sub)giant of any kind, the system was marked as \classE{}, and if \secondary{} is a white dwarf (WD), it was marked as \classW{} (cf. footnote\,\ref{fn_classes}). Neither is displayed in the figure. With that, we classified MS+MS systems (which are the interesting ones for a direct cluster comparison) based on the CPD positions of the components, whereas we classified all other systems based on the CMD position of the components.

    Table\,\ref{tab_binaries_age_bin} gives an overview of the numbers of wide binary systems that fall into all the aforementioned classes. The wide binary systems are also resolved by age group in the table. Table\,\ref{tab_period_sample} correspondingly includes a column that lists the class assigned to each wide binary system. Following our definition of the classes, all systems labeled as \classS{} or \classF{} agree with the open clusters (except those in age group\,7, which are beyond); those labeled \classC{} are contradictory. Figure\,\ref{fig_sample_ages} displays the distribution of wide binaries separated into the individual age groups\,1\,--\,7 and also distinguishes them by class. In Sect.\,\ref{sec_discussion}, we explore the patterns and distributions that emerged from the grouping and classification in detail. The commonalities are illustrated in Sect.\,\ref{sec_agreeing_systems}, whereas the disparities are investigated in Sect.\,\ref{sec_outlier}. Age group\,7 and classes \classE{} and \classW{} are discussed separately in Sect.\,\ref{sec_other_conclusions}.

\begin{sidewaysfigure*}
    \includegraphics[width=\linewidth]{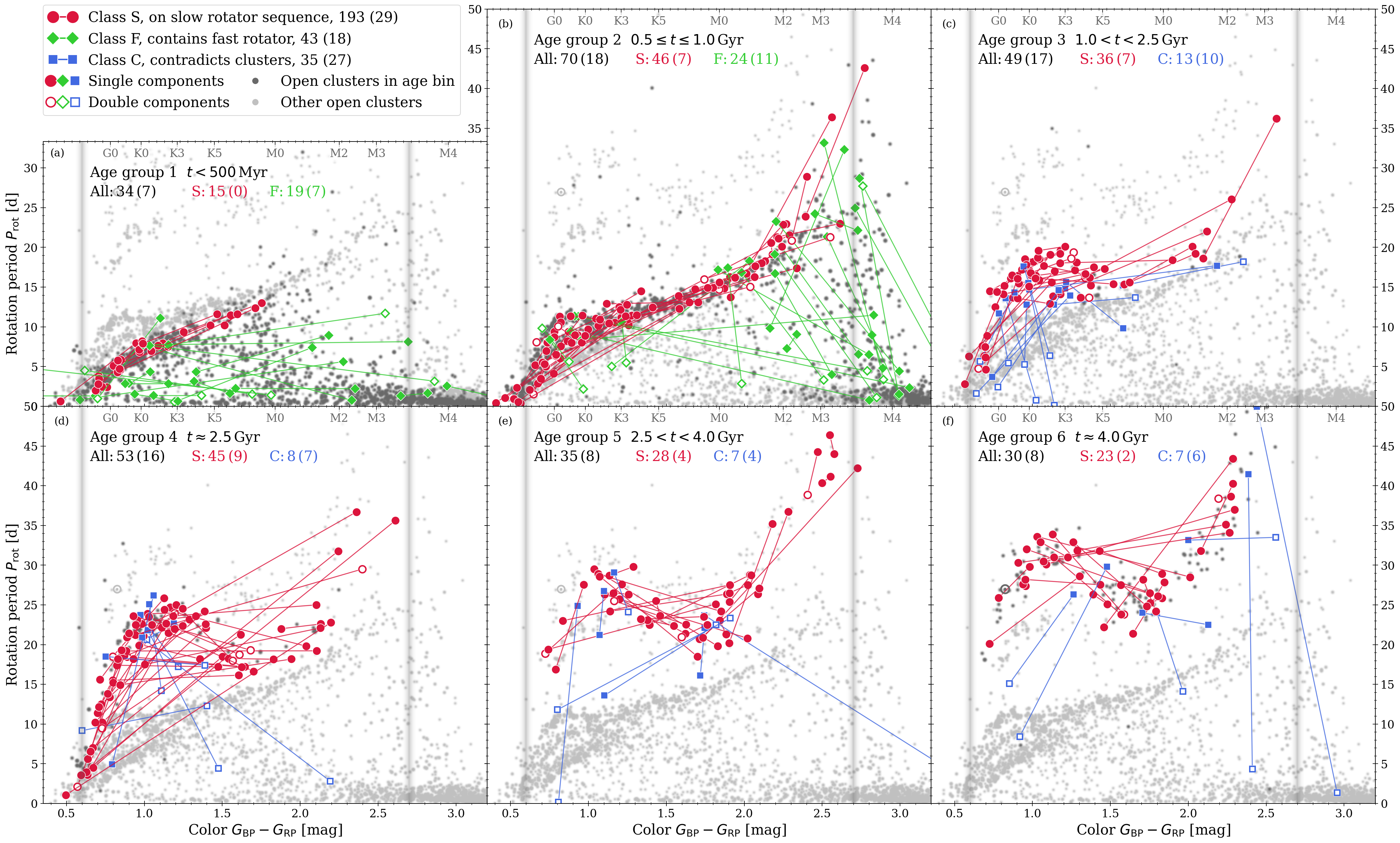}
    \caption{
        Color-period diagrams for six age groups shown in order of age {(left to right, top to bottom)}. The gray dots are open cluster stars, and cluster stars of the same age group are highlighted. Components of wide binaries are connected by lines and overplotted. Red circles indicate those agreeing with the slow rotator sequences (\classS), green squares are those connecting to a fast rotator (\classF), and blue boxes are those altogether disagreeing with the cluster behavior (\classC). Open symbols denote doubles (i.e., wide binary components identified as binaries themselves). Numbers in the upper-left corner of each panel enumerate the binaries in that age group, distinguished by class (see color coding). Numbers in parentheses indicate how many of these are hierarchical (i.e., where a component itself is a binary). The Sun is shown with its usual symbol. As in Fig.\,\ref{fig_sample_overview_cmdcpd}, vertical gray lines indicate the Kraft break and the fully convective boundary.
    }
    \label{fig_sample_ages}
\end{sidewaysfigure*}

\begin{table}[ht!]
    \caption{Population of wide binary systems within the individual age groups. The systems are also distinguished by class.}
    \centering
    \begin{tabular}{lcccccccc}
    \hline\hline
    \\[-0.8em]
Class                &          1 &          2 &          3 &          4 &          5 &          6 &  7\tablefootmark{b} &      Total\\ 
    \hline
    \\[-0.8em]
\classS{}            &         15 &         46 &         36 &         45 &         28 &         23 &          9 &        202\\ 
\classF{}            &         19 &         24 &         -- &         -- &         -- &         -- &         -- &         43\\ 
\classC{}            &          0 &          0 &         13 &          8 &          7 &          7 &          3 &         38\\ 
    \\[-0.6em]
MS\tablefootmark{a}  &         34 &         70 &         49 &         53 &         35 &         30 &         12 &        283\\ 
    \\[-0.6em]
\classE{}            &          0 &          1 &          4 &          7 &          3 &          4 &          2 &         21\\ 
\classW{}            &          4 &         17 &         13 &          8 &         15 &          8 &          3 &         68\\ 
    \\[-0.6em]
All                  &         38 &         88 &         66 &         68 &         53 &         42 &         17 &        372\\ 
    \hline
\end{tabular}  
    \tablefoot{
        \tablefoottext{a}{Class \emph{MS} is simply the sum of all wide binaries composed of two main sequence stars (i.e., excluding classes \classE \,and \classW).}
        \tablefoottext{b}{Age group\,7 includes three systems classified as \classC{} based on the analysis in Sect.\,\ref{sec_extrapolate}.}}
    \label{tab_binaries_age_bin}
\end{table}

\section{Discussion}\label{sec_discussion}

    In this section, we examine how our binary sample compares with the open clusters (Sect.\,\ref{sec_agreeing_systems}). We investigate outliers and identify probable causes for any deviations (Sect.\,\ref{sec_outlier}). Generally, a remarkable agreement between the wide binaries and open clusters is found. We follow up (Sect.\,\ref{sec_other_conclusions}) with a closer look at the wide binaries that either exceed the age range of the open clusters or for which one component is an evolved star.

\subsection{Systems in agreement with open clusters}\label{sec_agreeing_systems}

    We found that a vast majority (\agreeingbase{}\,\classS{} and \agreeingfast{}\,\classF) of the wide binary systems agree with the cluster sequences. Only a small fraction (\disagreeingwb{}\,\classC) does not. This means that \agreeingallbase{} out of \mswidebinariesbase{} total (MS+MS, excluding group\,7) wide binary systems agree well with the cluster predictions. They agree even to the extent that they follow the sinusoidal shapes of the sequences from the old clusters; they connect along and often across the peaks and troughs in the color dependence. The fact that the wide binaries agree with the clusters holds both for two very similar component stars (which tend to have very similar periods) and for two very different components (which may or may not have two very different periods). The above statements hold across all the age groups, from the youngest to 4\,Gyr (and likely beyond). The number of contradictory systems is relatively constant across the groups, except for age groups\,1 and 2, where there is not really one sequence and deviations are not apparent.

    Age\,group\,2 contains the largest number of wide binary systems in our sample. This is likely a consequence of its youth and the presence of the fast rotators, as both make it relatively easy to detect rotation periods. There is a clear transition from group 1 to group 2, seen both in open clusters and in wide binaries, of the fast rotator population moving redward. The remaining blueish ($G_\text{BP}-G_\text{RP}<2.0$) fast rotators in group\,2 are low in number, and all of them show signs of binarity.

    Age\,group\,3 lies in the gap region between clusters of ages 1\,Gyr and 2.5\,Gyr. It is notable that the wide binaries here seem to slot in perfectly between these two cluster populations, with a certain amount of (understandable) overlap only among the bluest stars, where distinguishing age groups is inherently difficult.

    Age\,group\,4 is the most striking of our sample. It contains wide binary systems that are directly comparable with the 2.5\,Gyr-old clusters NGC\,6819 and Ru\,147. Its stars range in color from close to the Kraft break to early M-dwarfs, in certain cases even within the same wide binary system. It is in this age group that we found the most impressive agreement between wide binaries and open clusters, as stars connect across large color ranges and fall on the same cluster sequence.

    Age\,group\,5 lies in the unpopulated region between clusters of ages 2.5\,Gyr and 4\,Gyr. As such, it resembles systems in group\,3. It is relatively sparsely populated compared to younger groups, and it shows a significant amount of scatter. However, this feature is partially by design. This group includes all the binaries that cannot be clearly associated with groups\,4 or 6.

    There is an apparent shift in population from the younger to the older groups regarding the color range populated by the corresponding wide binaries. With increasing age, the numbers begin to favor K-type stars instead of G-type stars. This is likely a consequence of the reduced variability shown by aging G-type stars, which makes them more difficult to detect in comparison with K and M spectral types.

    To summarize, we found that the vast majority of MS\,+\,MS wide binary systems agree with the individual cluster sequences and our corresponding age groups. The age groups represent a series that changes with age, albeit a rather rough one. In Fig.\,\ref{fig_colored_cpd_onecol}, we fitted the groups back together. We reduced the systems plotted to the \agreeingallbase{} agreeing ones (i.e., classes \classS{} and \classF{}) and add the nine{} systems from group\,7. The wide binaries form a clearly layered structure\footnote{Again, we note the parallels to stratigraphy in geology.}. Despite the significant scatter in certain regions of the CPD, the individual age groups are clearly separated in our wide binary sample.

\begin{figure}[ht!]
    \centering
    \includegraphics[width=8.8cm]{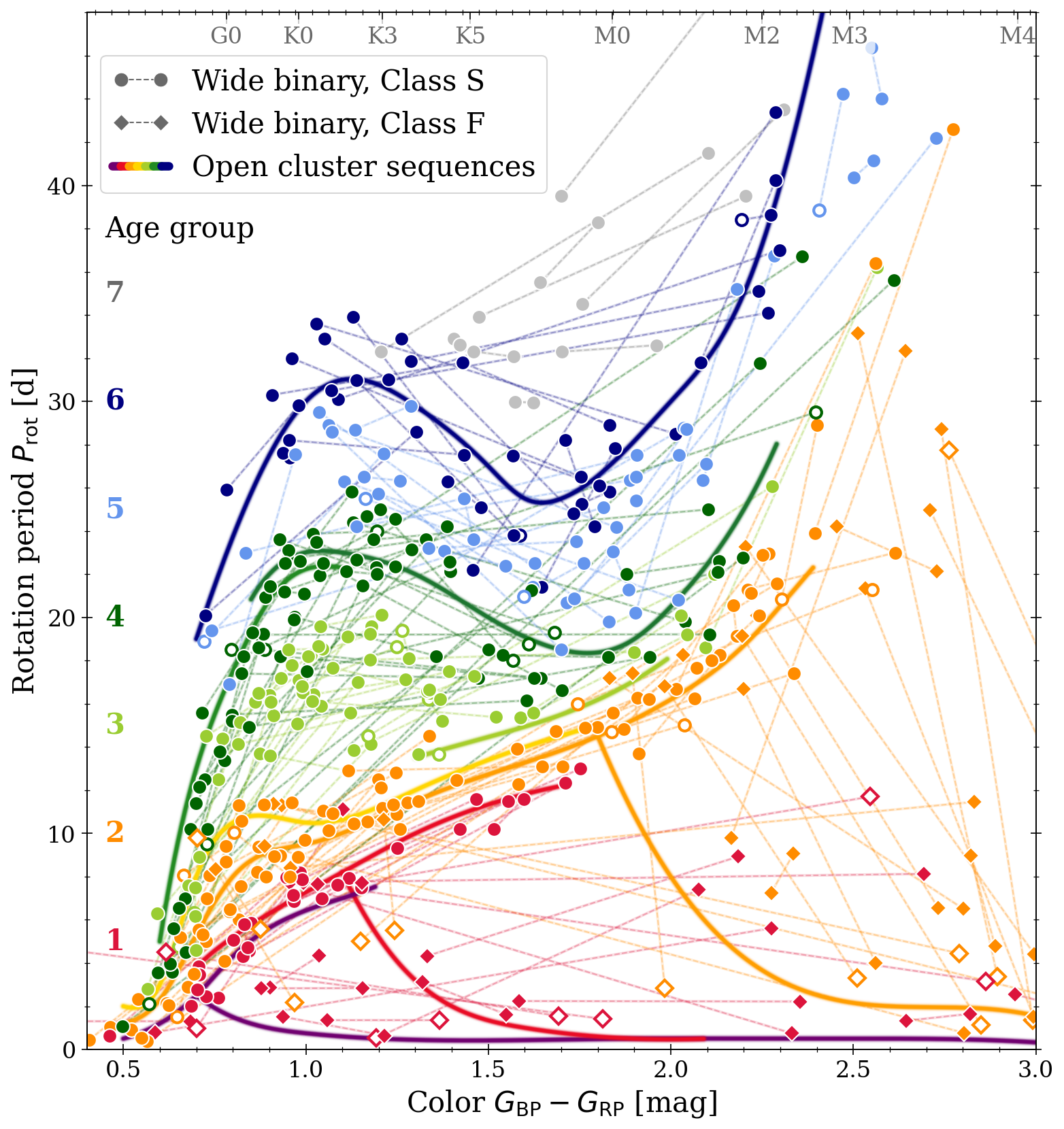}
    \caption{
        Color-period diagram showing the close agreement between the open cluster sequences and the wide binaries split into distinct age groups. The color coding follows the designation of the groups in Fig.\,\ref{fig_cluster_cpd} (see also "Age group" labeling). The open cluster sequences are overplotted with relevant colors. Only systems classified as \classS{} and \classF{} are shown. Again, open symbols denote stars identified as doubles.
    }
    \label{fig_colored_cpd_onecol}
\end{figure}

    A recent study by \cite{2023ApJ...947L...3B} recommends interpolating directly between clusters to estimate ages for stars in a slightly different but related way. They used the effective temperature $T_\mathrm{eff}$ and rotation period $P_\mathrm{rot}$ to estimate an age probability distribution. In Appendix\,\ref{sec_bouma}, we apply their software to our wide binary sample by estimating $T_\mathrm{eff}$ from the $G_\mathrm{BP}-G_\mathrm{RP}$ color of the stars. The calculated ages (including errors) and the temperature estimates are included in the electronic version of Table\,\ref{tab_period_sample}. We found reasonable agreement between the calculated ages and our group affiliations despite the fact that we might sort a given binary into another group based on either the joint behavior or preference for the leading companion in the system (cf. upper panels of Fig.\,\ref{fig_bouma_comparison}). In fact, the ages for the components of our systems lying consistently on the slow rotator sequence (i.e., class \classS{}) both correlate strongly ($r = 0.94$) with each other and have a relatively small dispersion $1\sigma = 0.28$\,Gyr (cf. lower panel of Appendix Fig. B1). This corroborates the assumed coevality of wide binary stars and demonstrates that age estimates based on a star's color and rotation period can provide consistent results when used carefully. To a good approximation, one may adopt the mean age for the class\,\classS{} wide binary components as the age of the system.

\subsection{Systems contradicting the open clusters}\label{sec_outlier}

    A big obstacle to definitive conclusions in prior work was the large number of systems that did not behave as expected. Therefore, we took a closer look at our outliers (i.e., those systems where components do not agree with what the open clusters define). We first highlight the quantity and distributions. In our total sample of \mswidebinariesbase{} wide binaries (MS\,+\,MS, excluding group\,7), we found \disagreeingwb{} pairs ($\sim$14\,\%) that do not agree with the open clusters. This fraction is already lower than what was observed in previous works (e.g., \cite{2008ApJ...687.1264M} and \citealt{2022arXiv221001137S}). Furthermore, unlike \cite{2022arXiv221001137S}, we do not see an increase in disagreement for the (presumably) older systems. We believe that this difference is a consequence of the nature of the comparison (i.e., comparing observations to observations rather than to models) because the models get worse for older ages (see their Fig.\,2, for example).

\subsubsection{Hierarchical systems}

    \citetalias{2023AnA...672A.159G} has shown that at the age of M\,67, it is exceedingly more likely than not that a star that shows signs of binarity (even only photometric binarity) is not suitable for gyrochronology, as it is likely to exhibit a contradictory rotation period in the sense that the rotation period lies recognizably distant from the single-star cluster sequence in the CPD. Interacting binaries transfer angular momentum from the binary orbit to stellar rotation via tidal interactions. Thus, they are in a sense rejuvenated and rotate faster than their compatriot single stars. These rejuvenated stars are not useful for gyrochronology, whether for its calibration or for its application. \citetalias{2023AnA...672A.159G} concluded that gyrochronology breaks down almost entirely for stars like this, and in their sample, only about 20\% of the stars with signs of binarity behave like single stars.

    This means that signs of binarity are a very good indicator that a star's rotation rate differs from the normal spindown (typically too fast). Conversely, we can say that if a star shows a rotation rate that does not fit its age group and we find signs of binarity for it, it is very likely (even more so the older a system is) that this binarity is responsible for that misfit.

    All this, of course, specifically refers to close binarity. Systems as distant as our wide binary components are not affected in this way. However, the individual components themselves may be much closer binary systems as well. These doubles are subject to the consideration above and such a wide binary system is called hierarchical.

    After constructing our sample, we queried every star in the \citetalias{2000AnAS..143....9W} database\footnote{\url{simbad.u-strasbg.fr/simbad}} and marked it as binary if it was listed there as such (e.g., as spectroscopic or eclipsing binary). We complemented this with an annotation if \citetalias{2022arXiv220800211G} photometry suggests a photometric binary. We found that \wballbinaries{} of our MS\,+\,MS wide binary systems are hierarchical (i.e., contain a double). Table.\,\ref{tab_period_sample} flags hierarchical wide binaries and individual stars suspected as doubles accordingly, and we highlight them in the relevant figures.

    Looking back at our outliers, we found that \outlierbinaries{} of the \disagreeingwb{} systems are hierarchical. They are displayed in Fig.\,\ref{fig_sample_outliers}. In agreement with our hypothesis of rejuvenation, we found all but one hierarchical systems to exist in a configuration where the double component exhibits rotation that is too fast for the age predicted by the non-binary component.

\begin{figure}[ht!]
    \centering
    \includegraphics[width=8.8cm]{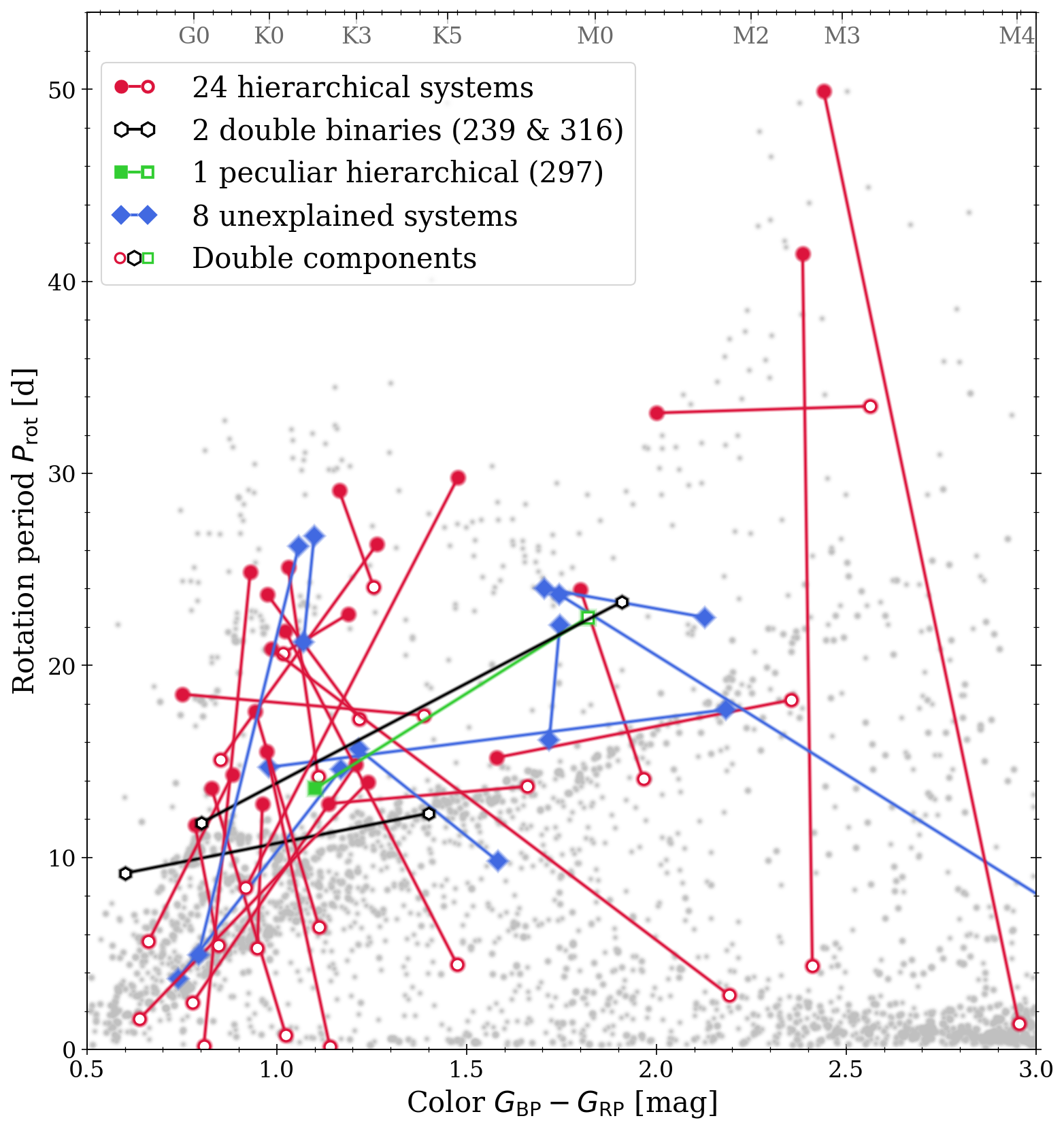}    
    \caption{
        Color-period diagram for the wide binary outliers. Open symbols indicate doubles, those wide binary components with binarity signs themselves (i.e., the wide binary system is hierarchical). Such hierarchical systems are displayed in red, while wide binaries in blue indicate those where we could not identify an obvious reason for not following the cluster sequences. Binary \hierarchicaloutlier{} (shown in green) is hierarchical, but it is the only system in our sample where the hierarchical component is not the one with a younger gyro age. 
    }
    \label{fig_sample_outliers}
\end{figure}

    The only exception to this pattern is binary\,\hierarchicaloutlier,{} whose secondary is \object{KIC 2442084}, an eclipsing binary (K-star + K-star) in a highly eccentric, relatively long-periodic\footnote{We mean "long" as compared with the expected rotation period.} orbit \cite[$e=0.599$, $P_\mathrm{orbit}=47.9$\,d,][]{2017AJ....154..105K}. The high eccentricity of this long-period orbital configuration is likely the reason for the inverted rotation in the wide binary, as the slow periastron passage is prone to decelerate a faster rotating star.

    It is also the case that a certain number of hierarchical systems lie uncontroversially on the rotational sequences we defined. Out of the \agreeingbase{} binaries in age groups\,1\,--\,6, \wbslowbinaries{}  in the \classS{} class and \wbfastbinaries{} in the \classF{} class are hierarchical. However, this fraction is much smaller ($\lesssim 20$\,\% of the systems per group, cf. Fig.\,\ref{fig_sample_ages}) than the $\outlierbinaries{}/\disagreeingwb{} \approx 76\%$ in class \classC{}, which is more along the lines of what was found by \citetalias{2023AnA...672A.159G}. We further note that the hypothesis of \citetalias{2023AnA...672A.159G} that the impact of binarity increases with age finds its confirmation here. While the fraction of hierarchical systems in our wide binary sample is relatively constant throughout the age groups, the fraction of those that do not agree with the cluster sequences increases with age (cf. counts listed in Fig.\,\ref{fig_sample_ages}: none in groups\,1 and 2, about half in groups\,3\,--\,5, and nearly all in group\,6). However, we stress that our sample only allows a conclusion based on relatively low number statistics in this regard.

    Our sample also contains two systems in which both stars are identified as binaries. Generally speaking, their position as a whole in a CPD (cf. black symbols in Fig.\,\ref{fig_sample_outliers}) is likely meaningless. Any assigned age group in such a case has to be viewed with commensurate caution.

    This assessment brought us to the following conclusions regarding close binarity. There is likely a (complicatedly shaped) boundary beyond which a close binary system remains rotationally unremarkable (i.e., both components evolve without interactions). This boundary likely depends on the orbital configuration (separation, component masses, eccentricity, etc.) but may not be such that we encounter multiple unresolved systems (i.e., likely relatively close) that are inconspicuous. Future efforts will have to explore this boundary, on samples like ours for example, in order to identify the differences between the affected and unaffected systems. One point, however, is already abundantly clear: Whether the rotation periods of close binary components will be affected is likely already determined during the system's formation. Gravitational encounters during a system's lifetime tend to drive close systems closer and wide system wider \citep[often summed up in the distinction between hard and soft systems; see e.g.,][]{1987gady.book.....B}. What remains is to evaluate the amplitude of the effect that angular momentum exchange has and to identify the boundary beyond which it loses significance. Until such understanding is attained, close binaries (and hierarchical systems like we discuss here) will remain inconclusive in the pursuit of exploring the rotational evolution of stars. And of course, all of this still leaves out two more aspects regarding unresolved binaries: The (combined) color may cease to be a valid proxy for the mass, and the origin of an observed rotational signal is somewhat ambiguous.

\subsubsection{Other outliers}

    After removing the hierarchical systems, there were nine outlier wide binaries remaining. We were unable to identify an obvious source for their contradictory behavior. However, there are a range of possible explanations for their apparent non-conformity.
    
    There can still be undetected multiplicity. This is likely the cause for most of the remaining deviations, especially for those wide binaries where one star rotates much faster than is expected from the other. An undiscovered planetary system (especially one hosting a Hot Jupiter) could also potentially be responsible.
    
    It is also possible that the wide binary is not actually genuine but merely a chance alignment. As such, the wide binary components would not be coeval. This is improbable here because of the stringent selection criteria in the source sample of \citetalias{2021MNRAS.506.2269E}. Other possible, albeit rather unlikely, scenarios include peculiar reddening that invalidates the color as a good proxy for the mass and angular momentum transfer from chance encounters.
    
    Independent of whether any of the mentioned arguments can explain the nature of an outlier system, we argue that nine remaining outlier systems out of \mswidebinariesbase{} is a small enough fraction not to matter to the overall conclusion. It is certainly small enough not to undermine the core principles discussed in this work: the validity of gyrochronology for field stars and the coevality of wide binaries.

    While none of the explanations given above are conclusive evidence for the deviations observed, they are good indications despite being somewhat based on circumstantial evidence. It could be argued that we have not offered a definitive criterion to separate rotationally well-behaved and ill-behaved systems. After all, many systems of class \classS{} (rotationally completely consistent) are also hierarchical (as discussed above). Nevertheless, we contend that whereas the single-binary nomenclature constitutes a strict dichotomy, the extent to which the rotation rates of components in a binary are influenced by such features as separation and eccentricity, among others, could potentially lie on a continuum, with only sufficiently isolated components being free to evolve rotationally as single stars. Certainly, the contours ofthat required isolation have not been mapped out to date. This thought is supported by the fact that open clusters also contain binaries, and some of them agree while some of them disagree with the single-star sequence.

    In any case, the situation again highlights an important point: If one is to adopt a sample of stars for calibration or verification purposes, one cannot rely on the statistical dominance of well-behaved, non-pathological systems but must scrutinize the nature (and data) of each and every star that is considered.

\subsection{Systems outside the range of open clusters}\label{sec_other_conclusions}

    Having established that wide binaries occupying regions of the CPD covered by open clusters actually agree with the cluster sequences, we next investigated systems where one or both components have evolved beyond available open cluster ages or past the main sequence. We divided this analysis into systems where both components are on the main sequence, systems where one component is a giant or subgiant, and systems where one component is a WD, with each category treated in its own subsection.

\subsubsection{Systems beyond the cluster sequences} \label{sec_extrapolate}

    Unlike the situation with the intermediate age groups 3 and 5 where we could interpolate reasonably well between the cluster sequences to identify concordant and discordant behaviors, we could not easily decide whether the binaries in age group\,7 agree with the cluster predictions, given that those require extrapolation. As such, we did not classify them like the earlier cases. However, we could still compare them with age group\,6 (4\,Gyr) and examine their internal consistency. Fig.\,\ref{fig_very_old_ones} shows the presumably older binaries of our sample in relation to the cluster predictions.

\begin{figure}[ht!]
    \centering
    \includegraphics[width=8.8cm]{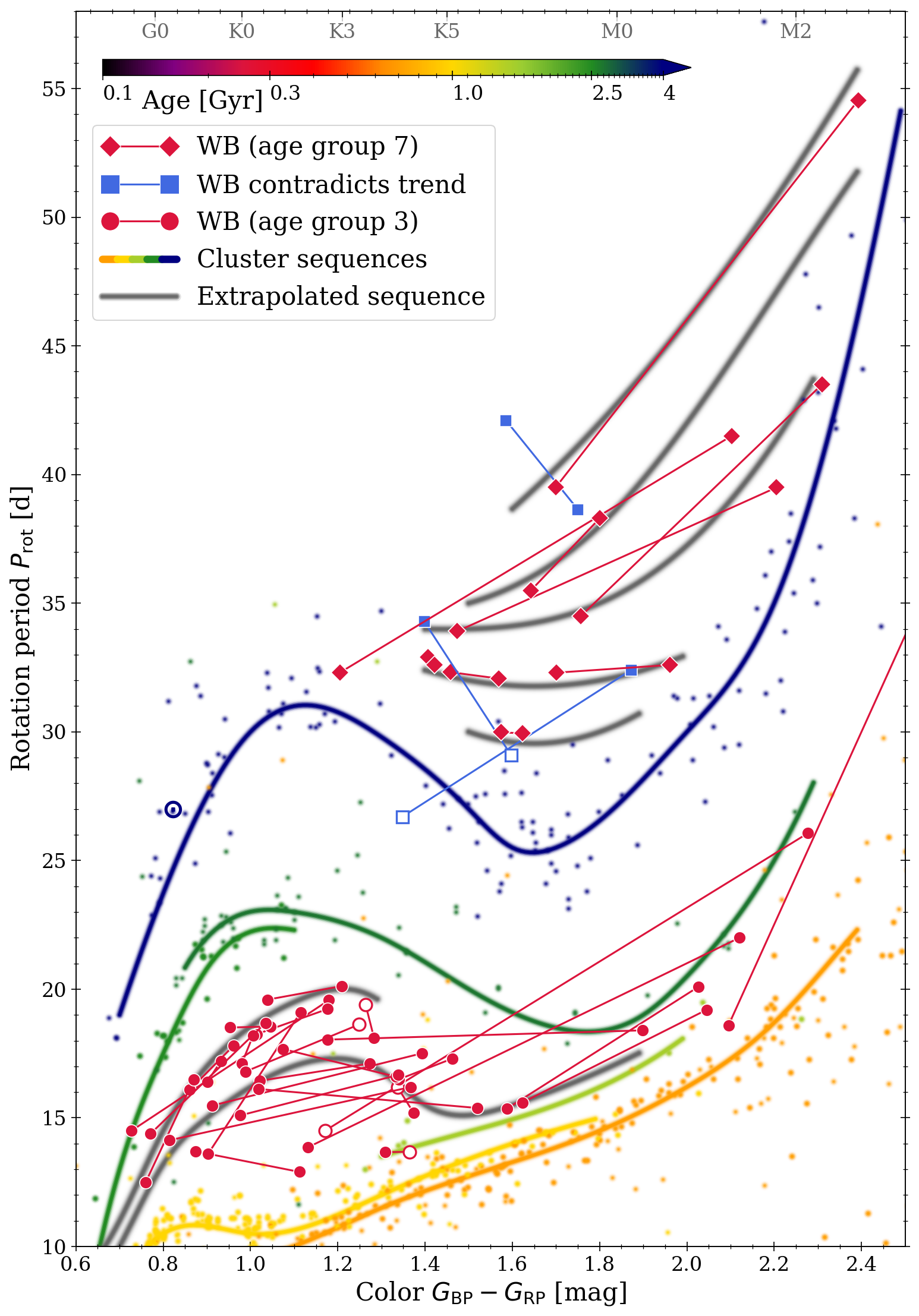}
    \caption{
        Color-period diagram showing a possible interpretation of the wide binaries older than 4\,Gyr and those with ages 1\,--\,2.5\,Gyr. The gray lines indicate a best guess extrapolation of the rotation sequences based on the wide binaries (red symbols) and adjacent cluster sequence (colored lines). The Sun is shown with its usual symbol.
    }
    \label{fig_very_old_ones}
\end{figure}

    We noticed the emergence of a certain trend from the distribution in group\,7 in Fig.\,\ref{fig_very_old_ones}. The explicit downturn in rotation period around $G_\mathrm{BP}-G_\mathrm{RP}=1.65$\,mag for 4\,Gyr seems to weaken slowly and eventually become a monotonic increase in period with color at even longer periods. In Fig.\,\ref{fig_very_old_ones}, we have drawn possible extrapolated sequences by eye to follow the stars in group\,7.

    Based on the extrapolation, we identified three contradictory wide binaries at older ages (highlighted in blue in the figure). We found that two of these are hierarchical, with the hierarchical component apparently rejuvenated.  Another one has a relatively large period error, and its mismatch could simply be a consequence of the generally large scatter in rotation periods observed in this long-period regime.

    One detail regarding our sample in age group\,7 is curious. That is, it only contains stars in the redder regions of the cool star mass range. The earliest-type star is $\sim$K3 ($G_\mathrm{BP}-G_\mathrm{RP}=1.2$\,mag). This is likely a consequence of the difficulty of detecting rotation signals in G-type stars rather than the absence of such stars altogether. We recall that the typical variability shown by the Sun would not be detectable by \emph{Kepler} at typical distances to stars in the \emph{Kepler} field. However, we do know about the existence of G-type stars with periods (and independent age estimates) in the relevant region, such as \object{HIP 102152} \citep[8\,Gyr-old solar twin with $P=35.7\,$d, \cite{2020MNRAS.tmpL..51L}; see also][]{2019MNRAS.485L..68L} and 94\,Aqr\,AB (see Sect.\,\ref{sec_famous_systems}).

    We attempted a similar artist's impression of the rotation sequences in age group\,3 ($1<t<2.5$\,Gyr, gray lines in Fig.\,\ref{fig_very_old_ones}). In this age range, the sinusoidal structure begins to emerge, and it appears to move with increasing age like a wave from blue to red. However, the maximum of this wave appears to be redder than what is observed in the open clusters at 2.5\,Gyr.

    A similar estimate for age group\,5 ($2.5<t<4$\,Gyr) was not possible, as the scatter of the wide binaries in this group is too great to be able to identify a suggestive picture. However, we note that the wide binaries are still consistent with a uniform (in a color-dependent sense) evolution between 2.5 and 4\,Gyr.

\subsubsection{Systems with a (sub)giant component}\label{sec_evolved}

    Our sample contains a number of systems (\binaryevo{}) that include (sub)giants as evolved components. The evolved star may or may not have a measured period itself. However, as the evolved component has changed its color due to its advanced state, expanded beyond its main sequence radius, and potentially experienced increased interaction with a planetary system, it is not expected to agree rotationally with the cluster predictions. One would need to invoke the mass directly for a meaningful comparison \citepalias[as was done for M\,67 by][see their Fig.\,16]{2023AnA...672A.159G}; this, however, goes beyond the scope of this work.

    We proceeded to investigate those systems in the same way as our main sample in that we considered the main sequence component as the \primary{} component and assigned it to an age group according to its period and color and then looked up the position of \secondary{}. However, in this case it is the position in a CMD. In Fig.\,\ref{fig_single_periods}, we compare \secondary{}'s position to an isochrone whose age is determined by the gyro age of \primary{}. For this comparison, we used (solar-metallicity) isochrones from the Padova and Trieste Stellar Evolutionary Code \cite[\citetalias{2012MNRAS.427..127B}\footnote{\url{stev.oapd.inaf.it/cgi-bin/cmd}},][]{2012MNRAS.427..127B,2014MNRAS.444.2525C,2015MNRAS.452.1068C}. As can be seen, the agreement is striking. All wide binaries except for one contain evolved stars whose CMD positions are consistent. The outlier (binary \targetoffevolved{}; cf. panels (a) and (e) of Fig.\,\ref{fig_single_periods}) is highlighted in red in the figure. We note that this outlier follows the same pattern as the other outliers (cf. Sect.\,\ref{sec_outlier}) in that the main sequence component appears rejuvenated (younger gyro age from faster rotation), as would be expected from the evolved component's CMD position\footnote{An argument could be made that we have listed only the half-period for the MS star.}. For the others, all deviations are small and can easily be attributed to metallicity, reddening, and some small amount of general scatter in the photometry. We also observed what was true for the main sample: This agreement holds throughout the age groups and for components of very different colors.

\begin{sidewaysfigure*}
    \centering
    \includegraphics[width=\linewidth]{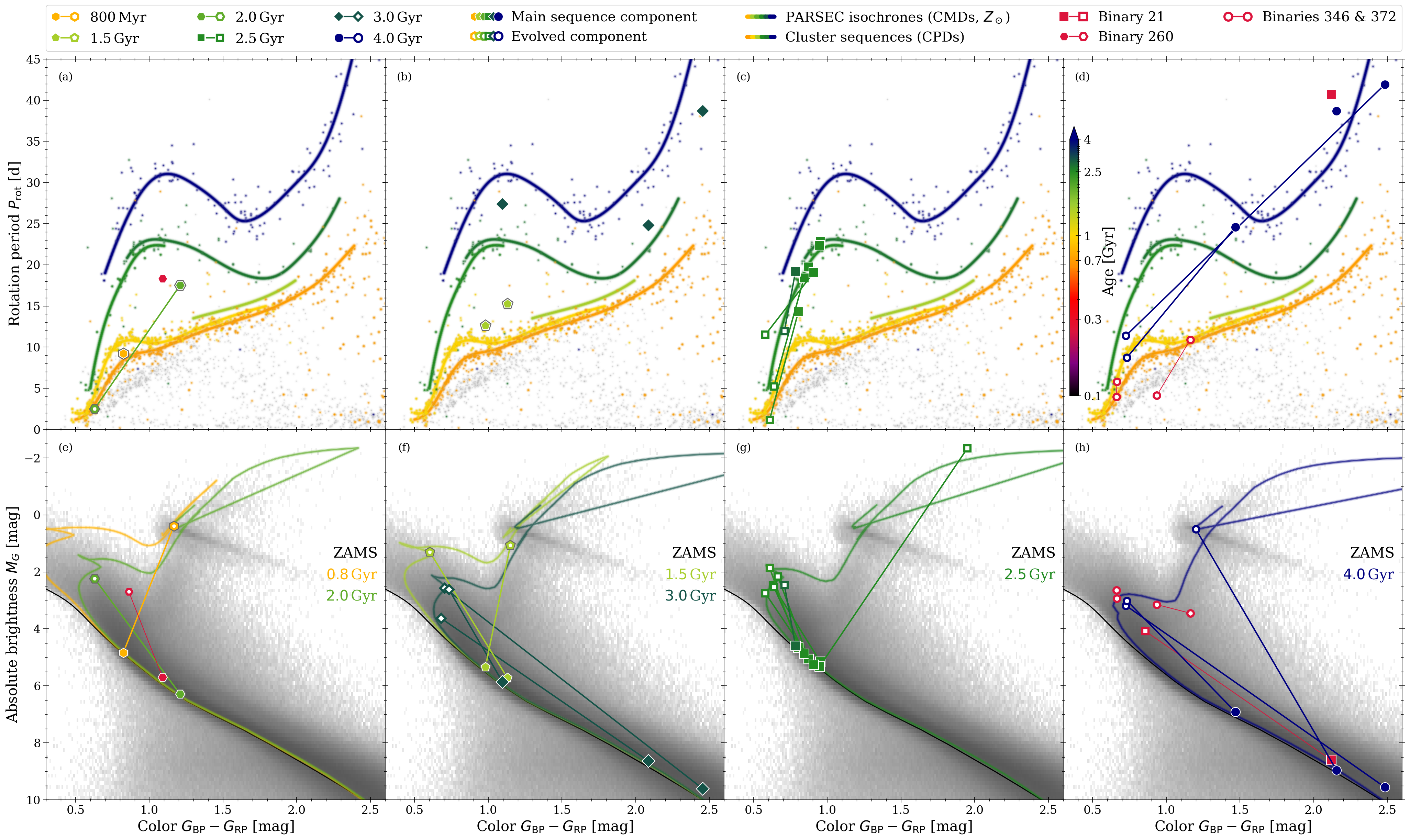}
    \caption{
        Color-period diagrams (panels a\,--\,d) and CMDs (panels e\,--\,h) for wide binaries  containing an MS (filled symbols) and evolved star (open symbols) combination. The CPDs show the wide binaries in relation to the open cluster sample. We recall that not all evolved components have known rotation periods. For those without, only an unconnected MS component appears in the CPD. The CMDs show the wide binaries (binned by age and color coded accordingly) against a background of the \emph{Gaia DR3} parameters for the full \cite{2021MNRAS.506.2269E} sample (2.6M stars). The isochrones (\citetalias{2012MNRAS.427..127B}) were chosen to match the colors of the components of each binary. Four binaries are highlighted in red and individually identified in the legend. These binaries include two wide binaries composed of two evolved stars (panel e), one that does not have a consistent CMD position (panel b), and one that is arguably older than 4\,Gyr. The separation of the CMD into four panels and the selection of the individual ages shown in them is for visibility purposes only. The CMD positions of binaries 21 and 346 suggest an age of $\approx 7$\,Gyr; however, no corresponding isochrone is shown for visibility reasons.
    }
    \label{fig_single_periods}
\end{sidewaysfigure*}

    Our sample also contains two wide binaries (binaries \targetbothevolveda{} and \targetbothevolvedb{}; cf. panel (d) and (h) of Fig.\,\ref{fig_single_periods}) that each have two evolved stars with identified rotation periods. Both systems appear rather old and are, in a CMD context, consistent. However, their CPD positions are not helpful to our work here, and we only list them for completeness. Finally, binary 21 consistently appears to be older than even M\,67 in both the CPD and CMD (cf. panel (d) and (h) of Fig.\,\ref{fig_single_periods}).

\subsubsection{Systems with a white dwarf component}\label{sec_white_dwarfs}

    Our sample provides an additional \binarywd{} binaries that contain a white dwarf (WD). It is always the MS star that has a measured rotation period -- most WDs do not even have recorded light curves. While a comparison with WD ages is beyond the scope of this work, we can still inspect the systems for superficial consistency. We know that WDs enter their sequence bright and hot and then slowly cool down, becoming fainter and redder with age. Figure\,\ref{fig_white_dwarfs} generally shows the expected distribution, with younger WDs (based on the main sequence star's gyro age) being higher up and older ones further down the WD sequence.
    \begin{figure}[ht!]
        \centering
        \includegraphics[width=8.8cm]{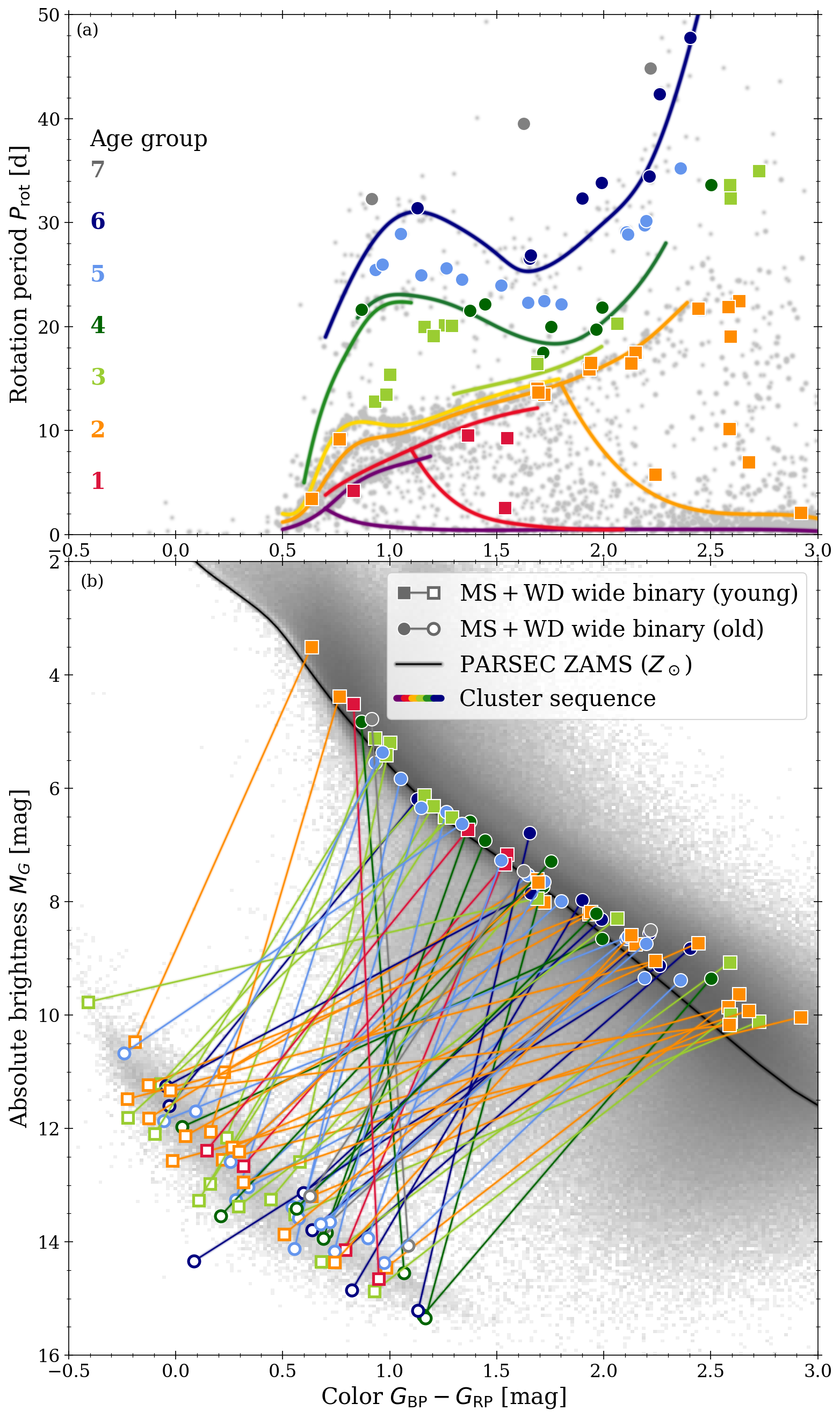}
        \caption{
            Color-period diagram and CMD for all wide binaries containing WDs. Panel (a): Color-period diagram of the MS components of the wide binaries color coded according to their age groups. Panel (b): Same sample of binaries shown in the CMD  but  connected to their WD companions (open symbols). The background gray scale distribution is of the 2.6 M stars of the \citetalias{2021MNRAS.506.2269E} sample. The MS stars with younger gyro ages (red-orange) show a slight preference to connect to the upper part of the WD sequence, while those with older gyro ages (green-blue) have a slight preference for the lower part of the WD sequence. Younger systems (age groups\,1\,--\,3) and older systems (age groups\,4\,--\,7) are shown with squares and circles, respectively.
        }
        \label{fig_white_dwarfs}
    \end{figure}
    Investigating individual deviants from this behavior exceeds the scope of this work. However, we do know that some of the MS stars are binaries themselves (not separately indicated here), a fact which may skew their rotation rate and age (generally toward younger ages). This could explain the odd, young WD further down the sequence. The older WDs further up the sequence do not contradict the expectations, as they may be WDs from lower-mass progenitors that have entered the WD phase later and have cooled down less.

\subsection{Revisiting well-known bright wide binaries}\label{sec_famous_systems}

    There are a number of familiar wide binaries that have been studied extensively in the past due to their brightness. In particular, these are some of the systems that have been considered in a rotational context by \cite{2007ApJ...669.1167B}, \cite{2008ApJ...687.1264M}, \cite{2014ApJ...780..159E}, and \cite{2022ApJ...930...36O}. We have compiled them together in Table\,\ref{tab_famous_wbs}. Rotation periods were taken directly from the above-mentioned works (where they agree reasonably well when they overlap; see the individual references therein). We supplemented the rotation data with \citetalias{2022arXiv220800211G} photometry and retrieved information regarding binarity from \citetalias{2000AnAS..143....9W}. These systems are not in our sample per se, as they were not in the \emph{Kepler} or \emph{K2} fields of view. An exception to this is 16\,Cyg, which is in our sample but was rejected because of cross contamination between the two very bright stars.

\begin{table}[ht!]
    \caption{Well-known bright wide binary systems.}
    \centering
    \begin{tabular}{lcccc}
    \hline\hline
Binary              & Component     &$P_\text{rot}$\tablefootmark{a} & $G_\mathrm{BP}-G_\mathrm{RP}$  & Notes\tablefootmark{b} \\ 
                    &               & [d]           & [mag]                          &      \\ 
    \hline
    \\[-0.8em]
\object{16 Cyg}     & A             & 23.8          & 0.81                           & Double    \\ 
                    & B             & 23.2          & 0.83                           &      \\ 
    \\[-0.8em]
\object{36 Oph}     & A             & 20.7          & 1.06                           &      \\ 
                    & B             & 21.1          & 1.06                           &      \\ 
                    & C             & 18.0          & 1.41                           &      \\ 
    \\[-0.8em]
\object{61 Cyg}     & A             & 35.0          & 1.46                           &      \\ 
                    & B             & 38.0          & 1.72                           &      \\ 
    \\[-0.8em]
\object{70 Oph}     & A             & 20.0          & 1.00                           & Double    \\ 
                    & B             & 34.0          & 1.49                           &      \\ 
    \\[-0.8em]
\object{94 Aqr}     & A             & 42.0          & 0.95                           & Double    \\ 
                    & B             & 43.0          & 1.06                           &      \\ 
    \\[-0.8em]
\object{alf Cen}    & A             & 28.0          & 0.84                           & Double    \\ 
                    & B             & 36.7          & 1.02                           &      \\ 
                    & C             & 83.0          & 3.80                           &      \\ 
    \\[-0.8em]
\object{ksi Boo}    & A             & 6.3           & 1.46                           &      \\ 
                    & B             & 11.9          & 1.52                           &      \\ 
    \hline
\end{tabular}
    \tablefoot{
        \tablefoottext{a}{Rotation periods were adopted from \cite{2007ApJ...669.1167B}, \cite{2008ApJ...687.1264M}, and references therein.}
        \tablefoottext{b}{"Double" indicates that a component has itself been identified as a binary.}}
    \label{tab_famous_wbs}
\end{table}

\begin{figure}[ht!]
    \centering
    \includegraphics[width=8.8cm]{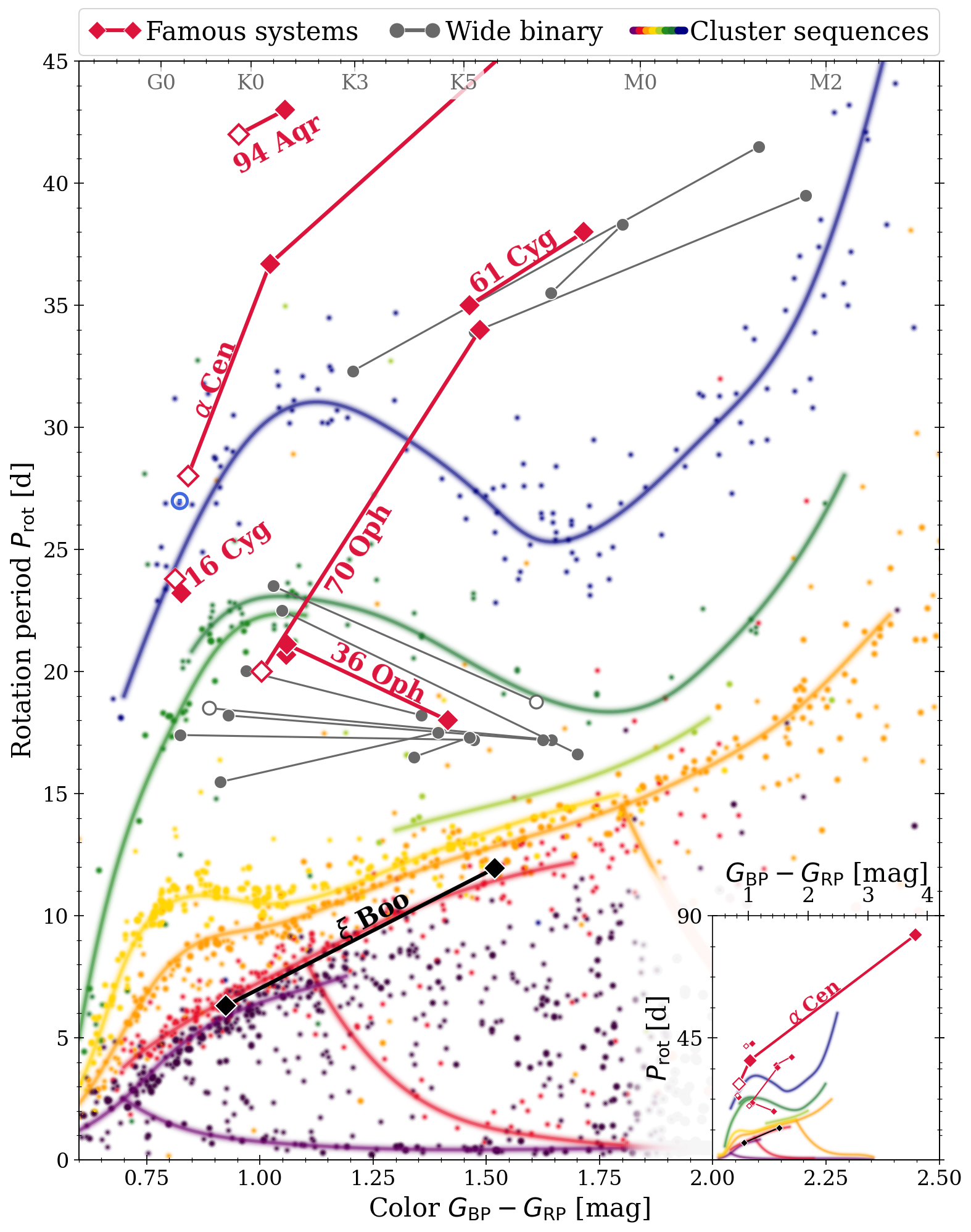}
    \caption{
        Color-period diagram for some well-known wide binaries in the literature. These are shown in red and labeled individually (see Sect.\,\ref{sec_famous_systems} and Table\,\ref{tab_famous_wbs}.) Open cluster sequences are overplotted together with a selection of wide binaries that follow comparable behaviors (where available) with the well-known ones. The inset shows the position of alf\,Cen\,C, which extends far beyond all other systems discussed in this work. As before, open symbols denote doubles (i.e., wide binary components that are themselves binaries).
    }
    \label{fig_famous_wbs}
\end{figure}

    We then compared the well-known systems with our sample and the open cluster predictions. Figure\,\ref{fig_famous_wbs} shows them in a combined CPD. We found significant changes in the picture since the initial publication of the above-mentioned work.

    ksi\,Boo is apparently the youngest among our sample of well-known systems. \cite{2007ApJ...669.1167B} reported gyro ages of 187 and 265\,Myr for the \emph{A} and \emph{B} components, respectively. \cite{2022AN....34320036S} have recently reported a lithium abundance for \object{ksi Boo B} that is consistent with that of stars in M\,34 (200\,Myr), corroborating the system's youth. We found that the rotation periods of both components in this system fall consistently on the slow rotator sequence of the 300\,Myr-old cluster NGC\,3532. 
            
    36\,Oph was anomalous in \cite{2007ApJ...669.1167B} because while the \emph{A} and \emph{B} components are almost identical (and thus automatically rotationally consistent), \emph{C} was found to rotate faster than expected (with gyro ages of $\sim$1.5\, for \emph{A} and \emph{B} and 0.5\,Gyr for  \emph{C}). \cite{2020ApJ...904..140C} pointed out that this system is rotationally parallel in the CPD with newly obtained data for Ruprecht\,147 and likely younger because the system is below the cluster sequence. We agree with this evaluation based on comparison with the same cluster and with similarly situated wide binaries in our sample (in groups\,3 and 4). Altogether, we found that all three stars of this system suggest a consistent age, likely around 2\,Gyr.
     
    16\,Cyg is unaltered with respect to prior work. Both stars are very similar in multiple ways, including rotation periods. As such, they are fully consistent with the M\,67 cluster and wide binaries of 4\,Gyr age, although it should be noted that the primary is somewhat evolved. Problems arise when one compares this assessment \citep[e.g.,][]{2016Natur.529..181V} with its asteroseismic age, which appears to be on the order of 7\,Gyr \citep[e.g.,][]{2020AnA...635A..26B,2022A&A...661A.143B}. However, since we are only concerned with its consistency with the open clusters and wide binaries, we did not engage any further with the system.
        
    70\,Oph was somewhat consistent with the old picture, pointing toward an age of 1.8\,Gyr, whereas at face value it is now inconsistent with what is seen for open clusters. The \emph{A} and \emph{B} components indicate very different rotational ages. Knowing that the \emph{A} component is a binary changes our perspective, and the system's age. Examining the system in the same way we did for all contradictory, hierarchical wide binaries, we found that the \emph{A} component rotates too fast compared to the age set by the \emph{B} component, which we estimate to be around 5\,--\,6\,Gyr. We note that the rotation period of 70\,Oph\,B was estimated from chromospheric activity. However, considering its similarity to 61\,Cyg\,A in terms of color and activity and the measured rotation period of 61\,Cyg\,A, it appears relatively reliable. We also note that the identification of 70\,Oph\,A as a binary is based on radial velocity variations \citep{2018A&A...619A..81H}. Those, however, indicate an orbit of approximately $88\,$years, making it more likely to be the \emph{A}\,-\,\emph{B} orbit rather than a hypothetical \emph{Aa}\,-\,\emph{Ab} orbit. 
        
    61\,Cyg is beyond the range populated by the open clusters in the CPD. Its components were always assumed to be consistent, but its estimated age of 6\,Gyr \citep{2008AnA...488..667K} did not match its early gyro age of 2\,Gyr \citep{2007ApJ...669.1167B}. \cite{2020ApJ...904..140C} evolved the cluster sequence of Ruprecht\,147 forward in time and found reasonably good agreement between its estimated age and its CPD position. From our current perspective, which also includes results for M\,67 and other wide binaries in the relevant region, we can confirm that the position of 61\,Cyg in the CPD appears to be fully consistent with an age of about 6\,Gyr.  
        
    Alf\,Cen is also beyond the range populated by the open clusters and our wide binaries. However, it gives an impression that the binarity of the \emph{A} component again impacts its rotation rate, making it rotate faster than it otherwise would. It was consistent with being about solar age in the (now superseded) descriptions used by \cite{2007ApJ...669.1167B} and \cite{2008ApJ...687.1264M}, as well as the forward prediction by \cite{2020ApJ...904..140C}. Given that we do not know the evolution (especially of G-type stars) beyond solar age, it very well may still be consistent. There is a recent debate in the community \cite[see e.g.,][and references therein]{2021NatAs...5..707H} about how spindown behaves for Sun-like stars older than solar age, but since we do not have clear indications from the open clusters, we do not engage in that discussion here. Alf\,Cen\,C, more commonly referred to as \object{Proxima Cen}, is far beyond any region populated by the clusters, both in terms of color as well as period.
        
    94\,Aqr is likewise beyond the range populated by the open clusters and the wide binaries. While its CPD position appears to be consistent, we note that the 94\,Aqr\,A is a Hertzsprung-gap star. Thus, the CPD agreement is coincidental. Evaluating the system using methods described in Sect.\,\ref{sec_evolved}, we found that the CMD position of \emph{A} suggests a younger age (3\,--\,4\,Gyr) than B's CPD position ($\gg 4$\,Gyr).
            
    To summarize, wide binaries that were formerly thought to be consistent (alf\,Cen, 70\,Oph) are now not. These wide binaries have also been found to be hierarchical systems, suggesting a possible resolution to their newly found inconsistencies with gyrochronology. Another system (36\,Oph) is not known to be hierarchical beyond the three resolved components, which are confirmed to be consistent with gyrochronology. Similarly, ksi\,Boo provides a consistent picture. The old age of 61\,Cyg that has sparked controversies in the past, thanks to its much younger gyro age, is seen to be fully consistent. With that, no unexplained inconsistencies remain.

\subsection{Metallicity effects}\label{sec_metallicity_effects}

    Theoretical considerations and observational indications suggest that a change in stellar metallicity changes the size of the convective zone, with consequent effects on the convective turnover timescale, the amount of differential rotation, and thus ultimately on the stellar dynamo and activity \citep[e.g.,][and references therein]{2018ApJ...852...46K}. However, the extent to which metallicity, by a similar route, influences stellar rotation and rotational evolution has yet to be understood. Rotationally investigated open clusters have largely been of solar metallicity. (In fact, the available observational evidence suggests that all nearby clusters are in a band of near-solar metallicity so that opportunities for related investigations are limited.) In principle, wide binaries, owing to their more diverse nature, offer a way past this. However, this means that we must first have an understanding of the stellar metallicity, which needs to be established individually for each binary.

    To do so, we obtained metallicities from two different sources: \citetalias{2022arXiv220800211G} and the \emph{TESS} Input Catalog \cite[TIC 8.2;][]{2019AJ....158..138S,2021arXiv210804778P}. \citetalias{2022arXiv220800211G} metallicities were measured from a small spectral range and provided as [Fe/H] values. Regardless of their uncertainties, they at least have the virtue of being uniformly derived. \citetalias{2022yCat.4039....0P} metallicities were compiled from a variety of spectroscopic surveys and are provided as [M/H] values.

\begin{figure}[ht!]
    \centering
    \includegraphics[width=8.8cm]{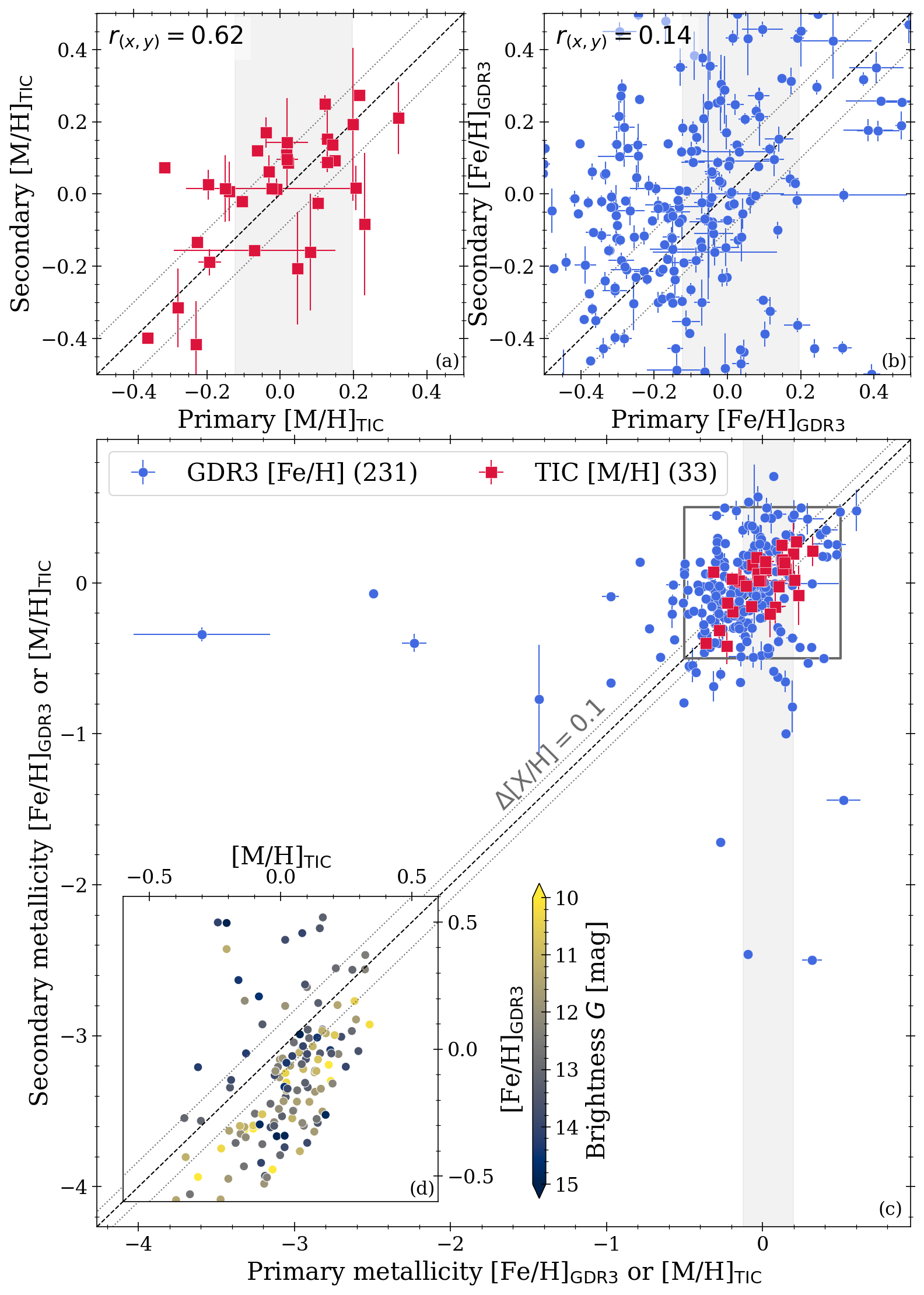}
    \caption{
        Comparison between the metallicities of the wide binary components of our wide binary sample, as extracted from \citetalias{2022arXiv220800211G} (blue, [Fe/H]) and \citetalias{2022yCat.4039....0P} (red, [M/H]). The highlighted region indicates the nominal metallicity range covered by the open cluster sample assembled in Table\,\ref{tab_cluster_compilation}. The gray box in panel (c) indicates the range plotted in panels (a) and (b). Panels (a) and (b) display Pearson's $r$ correlation coefficient \citep{1877Natur..15..492G,1895RSPS...58..240P} between the primary and secondary's metallicities. Panel (d) compares the [Fe/H] and [M/H] measurements for all our sample stars where both are available.
    }
    \label{fig_metallicity}
\end{figure}

    To begin this investigation, we checked each of these catalogs for consistency by comparing the metallicities of both wide binary components, when available. In principle, the metallicities should agree. However, as Fig.\,\ref{fig_metallicity} shows, the result is sobering. While the \binarieswticmetalicity{} wide binaries with \citetalias{2022yCat.4039....0P} values for the primary and secondary components at least display a correlation (coefficient $r = 0.65$), the \binarieswgaiametalicity{} wide binaries with \citetalias{2022arXiv220800211G} values barely even correlate with each other. It is almost certain that the low correlation arises from uncertainties in the metallicity measurements rather than in the association of the wide binary components. Consequently, we were unable to use the metallicities for the individual systems in a reliable way.

    However, we could make a more general observation. Although we may not know the individual metallicities precisely, it is likely that the binaries cover a much larger range. This is consistent with the fact that we drew our sample from both \emph{Kepler} and \emph{K2} data, which simultaneously probed very different regions in the Galaxy: the center and anti-center as well as within, above, and below the galactic plane, out to a distance of 1\,kpc. The rotation periods have already shown that the ages of the systems considered are very diverse, ranging from tens of millions of years to more than 4\,Gyr.

    Despite the diversity in the sample, the rotation period data does not show significant deviations from the cluster predictions. Our sample contains binaries in all age groups with components of very different masses, but nearly all agree with the open clusters. This suggests to us that changes in metallicity do not impact stellar rotational evolution in a significant way. It could, however, be the case that a change in metallicity affects stars of different mass equally, meaning in our case that both stars in the binary appear similarly displaced from their actual age in a CPD. In other words, a change in metallicity speeds up or slows down the spindown rate equally for stars of differing masses. This seems unlikely given that the spindown is strongly mass dependent.

    There is of course also the change in color with respect to a star of equal mass but different metallicity. Even a moderate variation in metallicity ($\Delta\text{[Fe/H]}\sim0.2$) causes significant change in the relation between color and mass for a main sequence star. Since we did not see a change in spindown in the color-period space, we would expect to see a change in the mass-period space. The explicit shape of this difference is unclear and beyond the scope of this work. It is, however, likely that the usage of color (or another photospheric parameter, such as temperature) somewhat diminishes the impact of metallicity changes with respect to a systematic spindown. Thus, it is a better approximation than the alternatives to state that stars exist on a single surface in age-rotation, period-color space. Further exploration of this will require a diverse sample of stars with well-constrained metallicities.

\section{Conclusions} \label{sec_conclusion}

    We investigated the rotation rates of stars in wide binaries by direct comparison with those in open clusters to explore the viability of using stellar rotation as an age indicator outside clusters, and perhaps even in field stars. To this end, we  constructed a sample of wide binaries whose components were observed by the \emph{Kepler} telescope during its primary and \emph{K2} missions. After scrutinizing the relevant data, adopting only verified rotation periods, and eliminating all those stars with periods that could not be verified, we created a clean sample of \allbinaries{} systems that shows remarkable agreement between the rotation rates of cluster stars and those of wide binary components.
    
    We find \agreeingallbase{} wide binaries where each contain two main sequence stars that agree right away with the cluster behavior. This includes wide binary pairs in all combinations of late-type main sequence stars ranging from the Kraft break to the boundary of full convection, and with ages ranging from tens of millions of years all the way to the age of M\,67 (4\,Gyr). The wide binaries in our sample follow the intricate mass dependence of the rotation rate in open clusters, especially those that were recently studied, including the flattening of the distribution around 1\,Gyr and the sinusoidal shape in older clusters.

    The sample also contains \evolvedms{} wide binaries that contain one component that has evolved beyond the main sequence into the subgiant and giant regions. For all but one of these, we find the CMD position of the evolved component to be consistent with the main sequence star's gyro age. An additional \binarywd{}  wide binaries contain a WD component, and generally speaking, their positions along the WD sequence are consistent as well.

    There are an additional nine{} diversely distributed wide binaries with dual MS components that are likely older than 4\,Gyr, permitting a rough extrapolation of stellar spindown and providing hints of the dependencies of rotation beyond ages covered by open clusters available to date. Lastly, assuming that our wide binary sample consists of a group of stars that has some diversity in composition, we conclude that the impact of metallicity on stellar spindown is likely small. 

    A minority of MS wide binary pairs (\disagreeingwb{} in number) are positioned in discordant rotation-age configurations. Three-fourths of these (\outlierbinaries) were shown to be hierarchical systems where angular momentum transfer has likely rejuvenated a component's rotation rate. This left us with only nine{} genuinely unexplained outliers. This is similar to the fraction of outliers in the cluster sample (e.g., the few blue, overly slow rotating stars in the \object{Praesepe} and \object{NGC 6811} clusters and the fast rotators in \object{M 67}).

    Altogether, these results suggest that the placement of well-characterized wide binary systems in the CPD is eminently comparable with open clusters of appropriate age. In fact, the components of clean wide binaries have the same rotational ages. Thus, wide binaries appear to be fundamentally compatible with the usage of gyrochronology. This compatibility likely extends to suitable single field dwarfs.

    The relative clarity of the results presented here is in contrast with the ambiguous results from prior work on wide binaries. We ascribe this difference to three factors:
        (1) Whereas prior work preferentially compared observations with models, ours is grounded in comparing observations with observations. Our ability to do so has been particularly enhanced by newly available observations, especially of the older Ruprecht\,147 and M\,67 clusters.
        (2) The sample of rotation periods admitted into our sample, although larger, is likely more exclusive, as it is informed by a greater scrutiny of the individual light curves and further recognition of the systematics and trending in both \emph{Kepler} and \emph{K2} data.
        (3) We performed a detailed investigation of outliers to identify disqualifying features, such as components being binaries themselves. We also identified systems with evolved components and found that they are generally in agreement with conclusions based on the rotation of the unevolved component.
    Rotation periods of evolved stars, although some are available, are currently unsuitable for gyrochronology.

    In conclusion, we find that the rotation of wide binary stars is demonstrably in agreement with open cluster stars, where the latter are available, and broadly in agreement with expectations for more evolved systems. Thus, our findings suggest that gyrochronology can likely be used to obtain ages for well-characterized single field stars. 

\begin{acknowledgements} 
    We are grateful to the anonymous referee for insightful comments and suggestions that helped to improve the quality of this paper.
    SAB gratefully acknowledges support from the Deutsche Forschungs Gemeinschaft (DFG) through project number STR645/7-1. 
    Some of the data presented in this paper were obtained from the Mikulski Archive for Space Telescopes (MAST). STScI is operated by the Association of Universities for Research in Astronomy, Inc., under NASA contract NAS5-26555. Support for MAST for non-HST data is provided by the NASA Office of Space Science via grant NNX09AF08G and by other grants and contracts. 
    This paper includes data collected by the \emph{Kepler} mission and obtained from the MAST data archive at the Space Telescope Science Institute (STScI). Funding for the \emph{Kepler} mission is provided by the NASA Science Mission Directorate. STScI is operated by the Association of Universities for Research in Astronomy, Inc., under NASA contract NAS 5–26555.
    This work has made use of data from the European Space Agency (ESA) mission {\it Gaia} (\url{cosmos.esa.int/gaia}), processed by the {\it Gaia} Data Processing and Analysis Consortium (DPAC, \url{cosmos.esa.int/web/gaia/dpac/consortium}). Funding for the DPAC has been provided by national institutions, in particular the institutions participating in the {\it Gaia} Multilateral Agreement. 
    This research has made use of the SIMBAD database, operated at CDS, Strasbourg, France. 
    Indications of dwarf stars' spectral types in this paper's CPDs and CMDs are based on ``A Modern Mean Dwarf Stellar Color and Effective Temperature Sequence'' \citet[][and continuously updated afterwards, Version 2022.04.16]{2013ApJS..208....9P}\footnote{\url{pas.rochester.edu/~emamajek/EEM_dwarf_UBVIJHK_colors_Teff.txt}}.
    This work made use of the \verb+python3+ implementations of the \verb+numpy+ \citep{harris2020array}, \verb+matplotlib+ \citep{Matplotlib}, \verb+scipy+ \citep{2020SciPy-NMeth}, \verb+numba+ \citep{10.1145/2833157.2833162}, and \verb+astropy+ \citep{2018AJ....156..123A,2022ApJ...935..167A} libraries.
\end{acknowledgements}


\bibliographystyle{aa}

\bibliography{lit}


\begin{appendix}

\section{Issues in the period sample}\label{sec_issues}

    To create our sample of wide binaries, we matched the \citet{2021MNRAS.506.2269E} catalog of wide binaries to \citetalias{2009yCat.5133....0K} and \citetalias{2017yCat.4034....0H} targets with known periods from \citetalias{2014ApJS..211...24M} and \citetalias{2020AnA...635A..43R}, respectively. However, there are a few issues regarding this matching, specifically, with the light curve data and with the derived periods. We illustrate these issues here in more detail.

\subsection{Light curve cross contamination}

    The spatial resolution of \citetalias{2022arXiv220800211G} (and also of \citetalias{2003yCat.2246....0C}) exceeds the spatial resolution of the \emph{Kepler} telescope CCDs. For a large number of sources with near neighbors, the point spread functions (PSFs) of the stars overlap and the aperture masks for each star include flux from the other. Thus, the light curves need to be checked for signs of this contamination.

\begin{figure}[ht!]
    \centering
    \includegraphics[width=8.8cm]{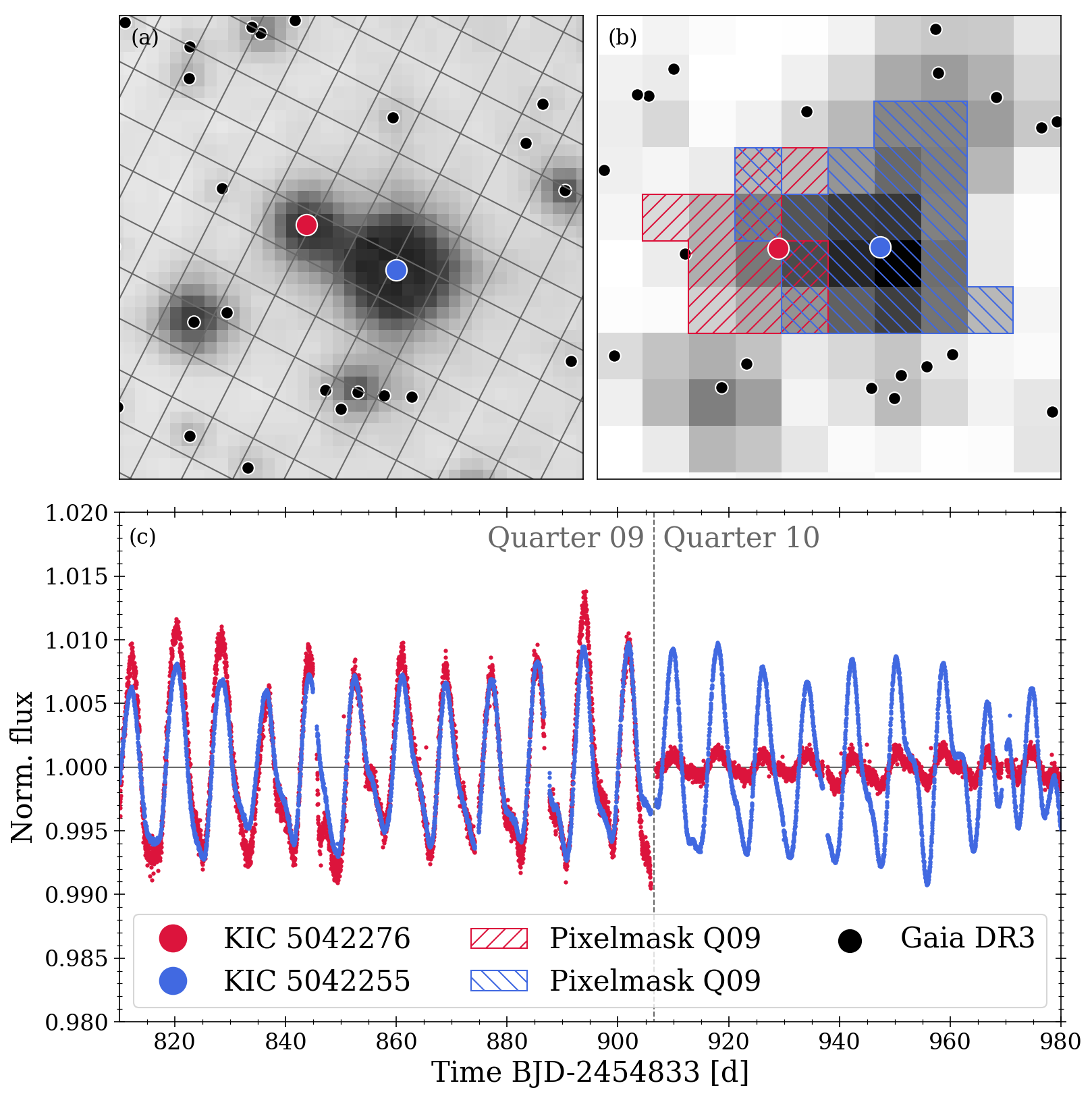}
    \caption{
        Example of light curve contamination. Panel (a): DSS2 (red) image around one of our wide binaries (\citetalias{2009yCat.5133....0K}; Ids given in the legend of panel c). The positions of the two wide binary stars (red and blue dots) are indicated, along with those of other sources recognized by \citetalias{2022arXiv220800211G}. The overlaid grid shows the \emph{Kepler} pixels that sample the region. Panel (b): Cutout from a \emph{Kepler} full frame image of the same region (with a somewhat similar orientation and size). The field of view in panels (a) and (b) is about $40\arcsec\times40\arcsec$. The overplotted hatched areas indicate the pixelmask for Quarter\,9 as reported in the auxiliary data to the light curves. Panel (c): Light curves of the component stars. They are focused on the end of Quarter\,9 and the beginning of Quarter\,10. The similarity in both behaviors is readily visible. The displayed and similar systems were discarded from our sample.
    }
    \label{fig_lightcurve_contamination}
\end{figure}

    We found \brokenbinaries{} wide binaries that suffer from this kind of cross contamination. Often times, one component of a wide binary affects the light curves of the other component (and vice versa). If we were to put the affected systems in a CPD, they would be rather obvious, as they have essentially the same rotation period and form horizontal structures. Figure\,\ref{fig_lightcurve_contamination} illustrates the light curves for one such system. It is clearly visible that both light curves exhibit the same behavior. The reported periods of the stars in \citetalias{2014ApJS..211...24M} are (as expected) essentially identical: 8.161\,d and 8.198\,d for \object{KIC 5042276} and \object{KIC 5042255}, respectively. The pixelmasks show the reason for this: The PSFs of the two components are too close to be separated. In this particular example, it is likely that the brighter star, \object{KIC 5042255}, is the origin of the observed variability and that the pixelmask of the secondary component records parts of the primary component's flux. The transition between Quarters\,9 and 10 involves a rotation of the telescope and a rather large change in the aperture mask for each star. The latter causes the changes between the recorded light curves for each star between the quarters. We note that the pixelmasks used in the \emph{Kepler} mission were created by a sophisticated algorithm that combines a stellar catalog with an estimate for the point spread function of a star to set an aperture. As the figure shows, this process is not (and cannot be) always fully successful. We rejected all \brokenbinaries{} wide binaries that showed signs of such contamination.

    An issue related to the cross contamination due to overlapping pixelmasks is one pixelmask capturing the flux of two stars. There are typically two cases: (1) The flux from a bright star located some distance away has been incorporated into the pixelmask for a fainter star. We observed this in \object{EPIC 220668834}, whose K-giant HD\,8412 companion \citep[$P_\mathrm{rot}=15.9\,$d,][based on ASAS data]{2012AcA....62...67K} created a false periodic signal observed in the light curve of the $\approx 11\,$mag fainter M-dwarf. (2) The pixelmask for a brighter star encompasses another star that is fainter but still bright enough to cast doubt on the origin of the observed signal. A good example is \object{EPIC 212694561} (and its faint companion \object{Gaia DR3 3625200307631332864}, both of which are a pair in our wide binary sample). Its light curve (see Fig.\,\ref{fig_mix_periods}) shows two distinct signals: one with about 14\,d and one with about 1.8\,d. \citetalias{2020AnA...635A..43R} reported a rotation period of  $P_\mathrm{rot}=13.8$\,d, the stronger signal, for \object{EPIC 212694561}. We could, in principle, make an educated guess regarding which period corresponds to which star. However, this runs the risk of prejudging the outcome, and thus we elected to proceed without introducing sample systems whose distribution would be informed by our prior assumptions. Accordingly, we rejected such systems from our sample.

\begin{figure}[ht!]
    \centering
    \includegraphics[width=8.8cm]{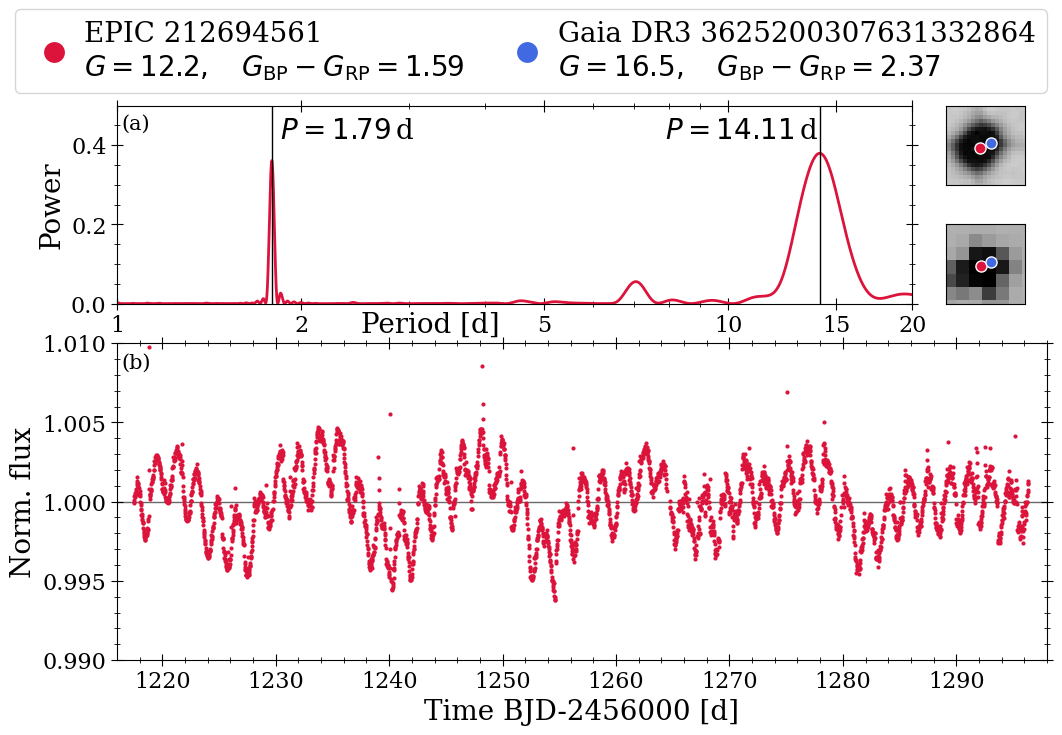}
    \caption{
        \emph{K2} light curve and power spectrum for \object{EPIC 212694561} showing the superposition of two periodic signals. Panel (a): Power spectra based on a Lomb-Scargle periodogram indicating the two periodic signals. Panel (b): Recorded light curve from \emph{K2}\,C06. The two small squares in the upper right show DSS2 (upper) and Kepler (lower) images around the recorded stars. The positions of the primary component, \object{EPIC 212694561}, and the secondary  component, \object{Gaia DR3 3625200307631332864}, are indicated. 
    }
    \label{fig_mix_periods}
\end{figure}

\subsection{Period identification}

    The problems caused by systematics and trends in \emph{Kepler} data (especially for \emph{K2}) have been described extensively elsewhere \citep[e.g.,][]{2014PASP..126..398H}. They impact period detections in \emph{Kepler} light curves, and as such we need to be aware of their existence and how they may impact the periods identified. Without going into too much detail, the important problem is that trends can mimic or obfuscate an intrinsic stellar signal. The problems intensify when the light curve only has a short baseline ($\lesssim 3 P_\mathrm{rot}$). Automated period finding algorithms (as employed by \citetalias{2014ApJS..211...24M} and \citetalias{2020AnA...635A..43R}) struggle to identify the right rotation period, especially for slower rotating stars, those with small amplitudes, multiple spots, and spot evolution. This was highlighted by \citetalias{2023AnA...672A.159G} (see their Sect.\,5.1) in an evaluation of rotation period work on M\,67.

    In space-based photometric time series data (such as what we obtained from \emph{Kepler}), which is both well-sampled and almost exactly regularly sampled, a rotation period is typically visually evident. However, it needs to be vetted against the adverse effects mentioned above. If carefully done, a manually set period is generally more reliable than an automatically derived one, albeit at the cost of being much more labor intensive. We strove for reliability in our sample and therefore manually re-derived the rotation periods for our sample stars based on an evaluation of the phase-folded light curves while accounting for spot evolution and considering the impact of data systematics (based on knowledge from light curves of similarly situated sources). Error ranges for the periods were also set manually by investigating how the phase-folded light curve behaves under changes of the assumed period.

\begin{figure}[h!]
    \centering
    \includegraphics[width=8.8cm]{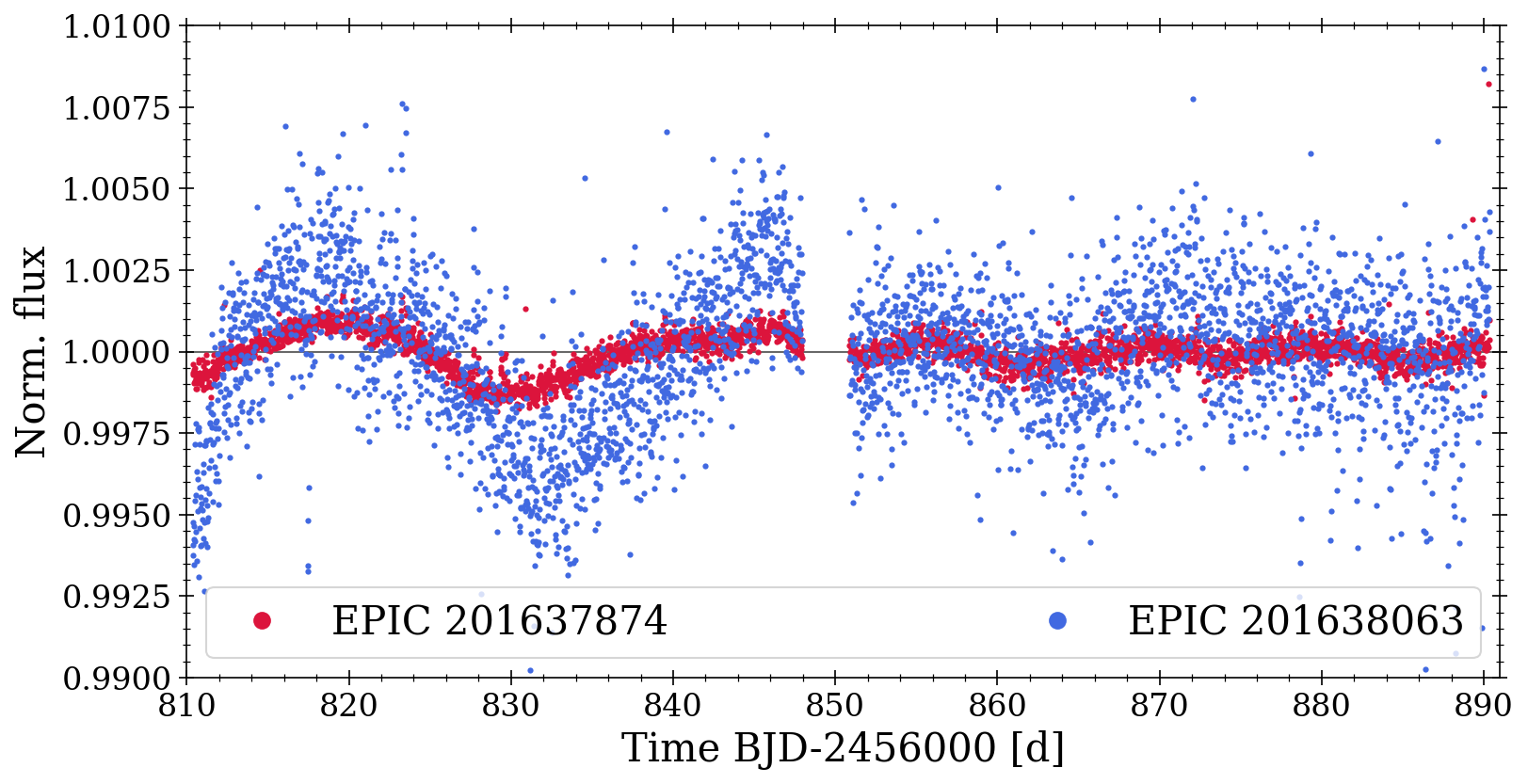}
    \caption{
        \emph{K2} light curves for two stars belonging to the same wide binary. Both stars show the same trending behavior, albeit with different relative amplitudes. The variation seen is apparently not from intrinsic variability of the star (for instance from a spot) but instead arises from trending; hence, the variation is an artifact.
    }
    \label{fig_lightcurve_trending}
\end{figure}

\begin{figure}[ht!]
    \centering
    \includegraphics[width=8.8cm]{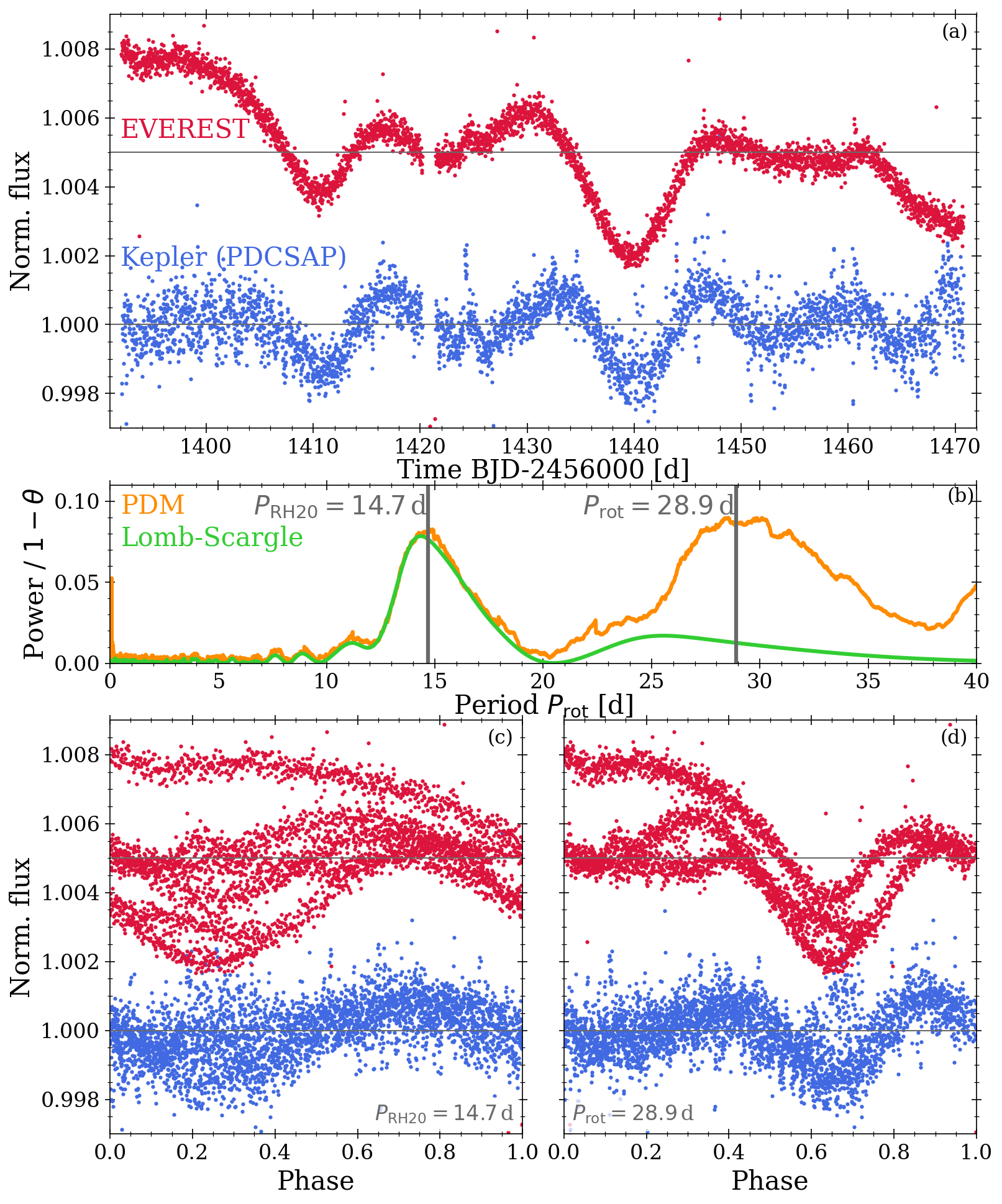}
    \caption{
        Light curve and period analysis for the sample star \object{EPIC 220332155}. Panel (a) shows the \emph{K2}\,C08 light curves obtained from the \emph{Kepler} archive (blue) and \citetalias{2016AJ....152..100L} (red). Both light curves show a clear double-dipping signal. The power spectra in panel (b) were calculated from a phase dispersion minimization (PDM; orange, displayed as $1-\theta$) and a Lomb-Scargle periodogram (green). The rotation periods identified by \citetalias{2020AnA...635A..43R} (14.7\,d) and us (28.9\,d) are both indicated. Panels (c) and (d) show phase-folded plots based on the \citetalias{2020AnA...635A..43R} period and ours, respectively. In panels (a), (c), and (d), we have shifted the \citetalias{2016AJ....152..100L} data vertically by 0.005 units for visibility reasons. Typically, trending is pronounced in \citetalias{2016AJ....152..100L} data; here, it is in the shape of a downward slope.
    }
    \label{fig_adjust_period1}
\end{figure}

    As expected, we encountered stars for which the reported periods by \citetalias{2014ApJS..211...24M} and \citetalias{2020AnA...635A..43R} do not reflect what we observe in their light curves. There are three types of problems contributing to this mismatch:
    
    (1) The reported period incorrectly identifies either somewhat random intrinsic variations of the light curve, systematic trending, or a combination of both as a periodic signal where there is none present. We rejected the corresponding \allmissingtoall{} stars from our sample. Figure\,\ref{fig_lightcurve_trending} shows an example of trending being identified as a rotation signal. Here, both stars are from the same wide binary, making the detection straight forward -- typically one needs to involve other stars in such assessments. Based on the figure, it is readily apparent that both stars exhibit the same (artificial) patterns, with both misidentified as periodic signals. And indeed, \citetalias{2020AnA...635A..43R} identified the same period of about 23\,d for both stars. We note that this similarity in the light curves is not an issue of contamination, as described above. Both stars are sufficiently far apart to be unproblematic with respect to the low spatial resolution of \emph{Kepler}.
    
    (2) The periodicity assigned was half of the true periodicity. \citet[][see also \citealt{2018ApJ...863..190B}]{2020AN....341..513T} have shown that longer period stars especially tend to exhibit multiple spot features in their light curves. These "double-dipping" stars can then be identified with only half of their real period, as the two spot signals are interpreted as the same. Figure\,\ref{fig_adjust_period1} shows an obvious example of such a case. We found \hpsstars{} such stars in our sample.
    
    (3) The star exhibits a clear periodic signal but the reported period is somewhat different from our preferred period. This occurred at times when the light curve exhibits strong signs of multiple spots and spot evolution that causes one spot to vanish and results in an erroneous association between variability features. Typically, the difference between the reported and the actual period is well within the error estimate. The error is large, due to being the result of a relatively bad fit, given the changes in the light curve behavior. Figure\,\ref{fig_adjust_period2} shows an obvious example of such a case. We chose to adopt our modified periods (and error) for these stars.
            
    Despite the described issues, we found that the vast majority of the stars (>90\,\%) are actually reported with the correct periods.

\begin{figure}[ht!]
    \centering
    \includegraphics[width=8.8cm]{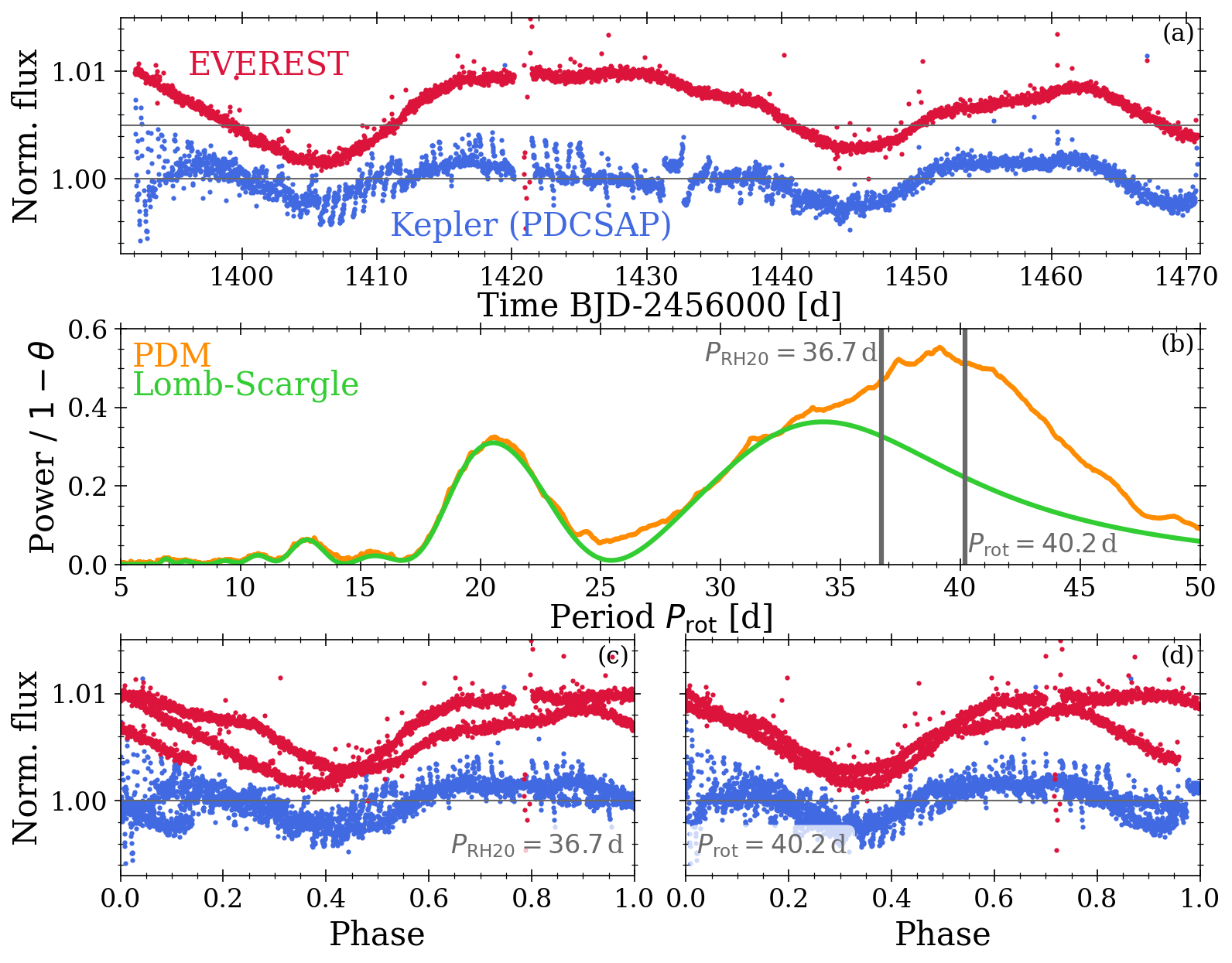}
    \caption{
        Same as Fig.\,\ref{fig_adjust_period1}, but for the sample star \object{EPIC 220652177}. Panel (a) shows the \emph{K2}\,C08 light curves obtained from the \emph{Kepler} archive (blue) and \citetalias{2016AJ....152..100L} (red). The light curves a clear double-dipping signal. The power spectra in panel (b) were calculated from a PDM (orange, displayed as $1-\theta$) and a Lomb-Scargle periodogram (green). The rotation periods identified by \citetalias{2020AnA...635A..43R} (36.7\,d) and us (40.2\,d) are both indicated. Panels (c) and (d) show phase-folded plots based on \citetalias{2020AnA...635A..43R} and our period, respectively. In (a), (c), and (d), we have shifted the \citetalias{2016AJ....152..100L} data vertically by 0.005 units for visibility reasons. 
    }
    \label{fig_adjust_period2}
\end{figure}

\subsection{Period comparison}

    Figure\,\ref{fig_period_comparison} shows a comparison between our final sample of periods and the values reported in \citetalias{2014ApJS..211...24M} and \citetalias{2020AnA...635A..43R}. As can be seen, the majority of the values are in agreement. However, we had to revise a significant number of periods, especially those for \emph{K2} stars. The changes are typically of the order of 10\,--\,20\%. There are \hpsstars{} stars for which we found that \citetalias{2020AnA...635A..43R} only reported half the period of a double-dipping star. In the one case where it appears that we had adopted the half-period of a \citetalias{2014ApJS..211...24M} star, it seemed that \citetalias{2014ApJS..211...24M} identified a period based on a large-scale systematic, while ignoring the weaker but more consistent signal throughout the rest of the light curve. For the other outliers where we identified $\approx 10\,$d periods, we were unable to understand what led \citetalias{2014ApJS..211...24M} to adopt  periods of only $\approx 1$\,d. 
    \begin{figure}[ht!]
        \centering
        \includegraphics[width=8.8cm]{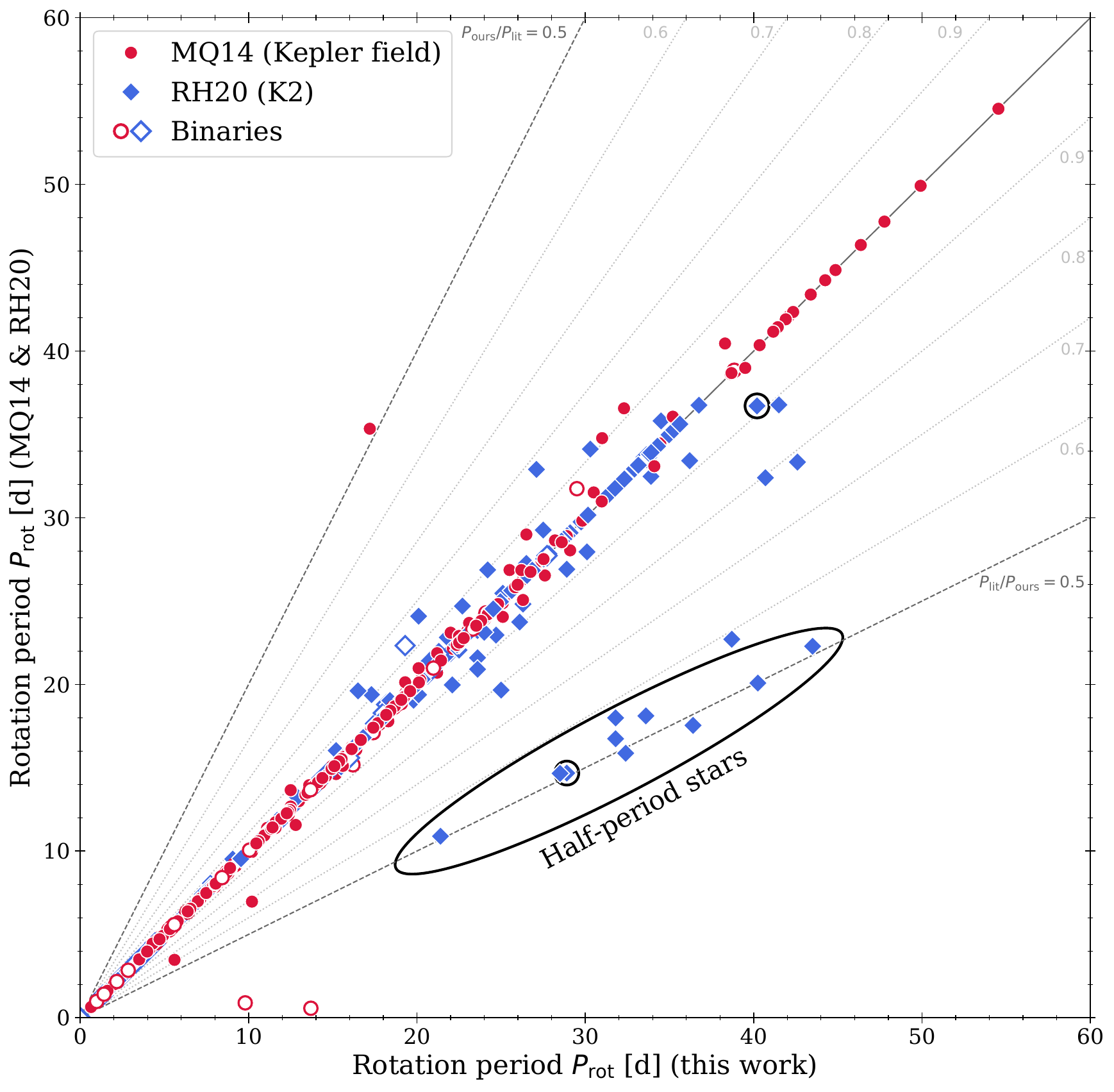}
        \caption{
            Comparison between the periods reported in \citetalias{2014ApJS..211...24M} (red) for the wide binaries in the \emph{Kepler} mission and in \citetalias{2020AnA...635A..43R} (blue) from \emph{K2} and our adopted and vetted periods. Overplotted gray lines indicate the size of the offsets between our periods and theirs, as labeled. Open symbols denote stars with signs of binarity. The black ellipse marks double-dipping stars that were detected with their half-period. The two highlighted stars are the ones displayed in Figs.\,\ref{fig_adjust_period1} and \ref{fig_adjust_period2}.
        }
        \label{fig_period_comparison}
    \end{figure}
    We also believe that the errors reported in \citetalias{2020AnA...635A..43R} for stars with slower rotation ($\gtrsim 20$\,d) are highly overestimated (e.g., $\approx 10$\,d for stars with $P_\mathrm{rot}=30$\,d).

\section{\texttt{gyro-interp} \texorpdfstring{\citep{2023ApJ...947L...3B}}{}}\label{sec_bouma}

    \citet[][\citetalias{2023ApJ...947L...3B} hereafter,]{2023ApJ...947L...3B} created an interpolation on the empirical sequences of selected open clusters using Bayesian statistics to estimate an age probability distribution for a star with a given effective temperature $T_\mathrm{eff}$ and rotation period $P_\mathrm{rot}$. In this section, we discuss the application of their algorithm (\texttt{gyro-interp}\footnote{\url{github.com/lgbouma/gyro-interp}}) to our wide binary sample. We proceeded as follows:
        (1) We selected all wide binaries from the age groups\,1\,--\,5 falling into the classes \classS{}, \classF{}, and \classC{}. 
        (2) For each star, we obtained $T_\mathrm{eff}$ from \citetalias{2022arXiv220800211G} $G_\mathrm{BP}-G_\mathrm{RP}$ via the empirical color-temperature relation from \citep[][see their Table\,4]{2020ApJ...904..140C}.
        (3) We assumed a flat $T_\mathrm{eff}$ error of 100\,K for each star for the calculation.
        (4) The calculation was executed on an age grid between $t=0$ and 4\,Gyr and with $\Delta t=0.01\,$Gyr.
    Not all stars can be processed this way. For example, the bluest and reddest stars of our sample are beyond the interpolation range of \texttt{gyro-interp}. We only obtained results for stars with $0.7 \leq G_\mathrm{BP}-G_\mathrm{RP} \leq 2.0$ (3800\,--\,6200\,K). Just as \citetalias{2023ApJ...947L...3B} did, the median of the distribution was adopted as the age $t_\text{\citetalias{2023ApJ...947L...3B}}$.

    In Fig.\,\ref{fig_bouma_comparison}, we illustrate the results of the calculation. As we did not have a precise age estimated from our procedure, we compared the calculated ages $t_\text{\citetalias{2023ApJ...947L...3B}}$ to the extent of the corresponding age groups. Panels (a)\, through\,(e) in the figure show histograms of the resulting age distributions distinguished by class. Each panel focuses on a different age group, and the histograms only show stars that we assigned to the particular (highlighted) group. Panel (f) shows a comparison between the wide binary components for all systems where we had an age estimate for both.

\begin{figure}[ht!]
    \centering
    \includegraphics[width=8.8cm]{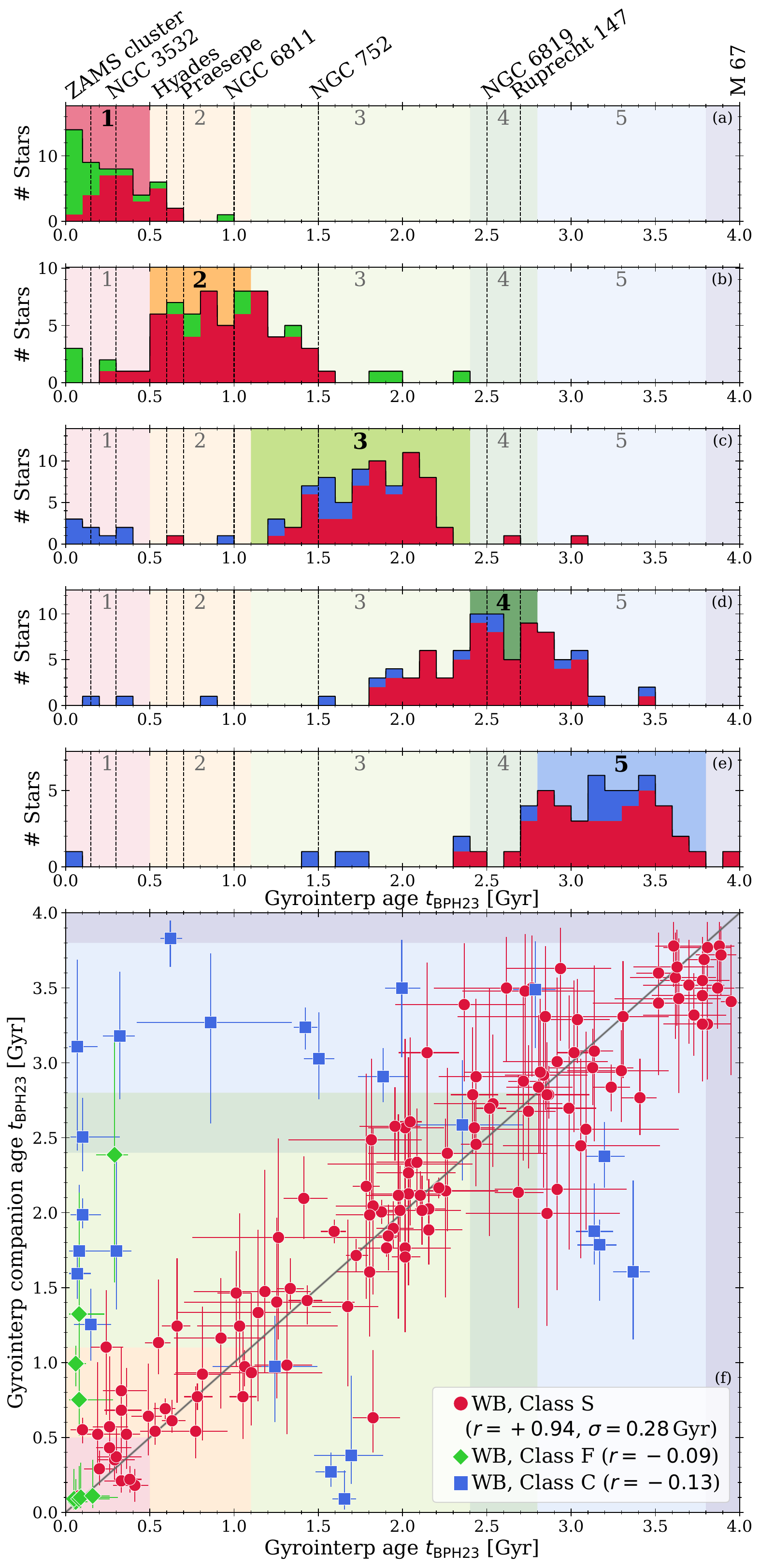}
    \caption{
        Comparison with ages from \texttt{gyro-interp}. Panels (a) to (e) show histograms of the ages sorted into our age groups\,1\,--\,5. The background color coding shows the rough extent of the individual groups, with the relevant one highlighted. Panel (f) shows the age of the wide binary component with a larger age uncertainty against the one with a smaller uncertainty. In all plots, the color coding refers to classification of the wide binary systems.
    }
    \label{fig_bouma_comparison}
\end{figure}

    Generally speaking, the agreement is good, although a significant number of stars (especially in groups\,2 and 4) lie beyond our nominal group limits. This is somewhat reflected in the age uncertainties, which are  not seen in the histograms. The largest discrepancies were in the bluest stars. This is a consequence of the strong mass dependence of the spindown in that region. Although \citetalias{2023ApJ...947L...3B} is cautious about age estimates beyond 2.6\,Gyr (as there are boundary effects from interpolation with only M\,67 as an older anchor point), we found that \texttt{gyro-interp} does a reasonable job for the stars in that age range. The best agreement was achieved for stars (and systems) that fall into the range (color- and rotation period-wise) that we mainly used to draw the \primary{} component in Sect.\,\ref{sec_analysis}. We note that \texttt{gyro-interp} appears to underestimate ages for fast rotators redder than the Sun. This hints at an underlying issue with the distinction between slow and fast rotator sequences at a given age and the sparsely populated gap between them (a non-trivial problem). The class \classC{} systems contain one star that rotates too fast for its assigned age, and that is reflected in the scatter in all plots.

    We consider \texttt{gyro-interp} to be a good first step toward obtaining model-independent rotational ages. Its current limitations arise mainly from the availability of open cluster data. We agree with \citetalias{2023ApJ...947L...3B} that future endeavors should focus on expanding the age range and closing gaps. A short-term improvement in the usability may be achieved by estimating ages directly using (dereddened) colors rather than $T_\mathrm{eff}$ values (e.g., by having \texttt{gyro-interp} calculate $T_\mathrm{eff}$ on the fly from provided colors). This would make it fully empirical and also more accessible, as photometric colors are typically far more extensively available than $T_\mathrm{eff}$.

\section{Sample table}

\begin{sidewaystable*}
    \caption{Excerpt of the wide binary sample. \label{tab_period_sample}}
    \begin{tabular}{lcccccccccccl}
    \hline\hline
    \\\\[-1.9em]
Binary     & Age                       & System                        & System          & Gaia DR3                  & 2MASS                & EPIC       & KIC        & $P_\mathrm{rot}$ & $P_\mathrm{err}$ & $G_\mathrm{BP}-G_\mathrm{RP}$  & $G$                            & Notes on  \\ 
ID         & group                     & Class\tablefootmark{a}    & Issue           &                           &            &                      &            & [d]             & [d]             & [mag]                          & [mag]                          & components\\ 
    \cline{9-10} 
    \\\\[-1.9em]
    \hline
\\[-0.6em]
37         & 1                         & \classF{}                 &                 & 2119651091691155712       & J18500021+4735070    & --         & 10386229   & 7.72       & 0.02       & 0.96                           & 11.8                           &           \\ 
           &                           &                           &                 & 2119667691742387456       & J18491928+4743299    & --         & 10515986   & 0.75       & 0.01       & 2.33                           & 15.5                           &           \\ 
\\[-0.6em]
68         & 4                         & \classC{}                 & hierarchical    & 2101369172560195840       & J19181006+4027333    & --         & 5267544    & 23.8       & 0.7        & 0.97                           & 15.1                           &           \\ 
           &                           &                           &                 & 2101369172560194816       & J19180995+4027289    & --         & 5267541    & 17.2       & 1.2        & 1.22                           & 15.3                           & Double    \\ 
\\[-0.6em]
133        & 1                         & \classF{}                 &                 & 2082192658985723264       & J20030145+4529233    & --         & 9118981    & 1.51       & 0.01       & 0.94                           & 13.7                           &           \\ 
           &                           &                           &                 & 2082191765632529664       & J20031048+4527271    & --         & 9119108    & 0.65       & 0.01       & 1.22                           & 14.9                           &           \\ 
\\[-0.6em]
148        & 2                         & \classF{}                 &                 & 2551656808041899136       & J00523879+0337462    & 220346755  & --         & 21.4       & 2.4        & 2.53                           & 15.5                           &           \\ 
           &                           &                           &                 & 2551656808041898880       & J00523729+0337516    & 220346833  & --         & 1.51       & 0.01       & 2.99                           & 16.6                           &           \\ 
\\[-0.6em]
156        & 4                         & \classC{}                 & hierarchical    & 2554578725832963840       & J00412128+0529399    & 220440269  & --         & 21.8       & 1.5        & 1.02                           & 13.4                           &           \\ 
           &                           &                           &                 & 2554578691473225600       & J00412782+0529419    & 220440299  & --         & 4.44       & 0.01       & 1.47                           & 14.9                           & Double    \\ 
\\[-0.6em]
183        & 4                         & \classS{}                 &                 & 2077301275348100352       & J19362795+4040518    & --         & 5456319    & 23.5       & 2.5        & 1.03                           & 14.6                           &           \\ 
           &                           &                           &                 & 2077300759952080896       & J19362346+4038341    & --         & 5456253    & 18.8       & 0.3        & 1.61                           & 15.8                           & Double    \\ 
\\[-0.6em]
186        & 4                         & \classS{}                 &                 & 2107342716152550272       & J18582249+4626591    & --         & 9696358    & 15.6       & 1.5        & 0.72                           & 11.7                           &           \\ 
           &                           &                           &                 & 2107336874997022464       & J18581799+4626119    & --         & 9696331    & 22.6       & 2.2        & 0.98                           & 14.3                           &           \\ 
\\[-0.6em]
189        & 4                         & \classC{}                 & hierarchical    & 2077605840068347904       & J19343327+4137114    & --         & 6290800    & 20.9       & 4.6        & 0.99                           & 13.3                           &           \\ 
           &                           &                           &                 & 2077605840068352768       & J19343395+4137235    & --         & 6290811    & 2.83       & 0.01       & 2.19                           & 15.4                           & Double    \\ 
\\[-0.6em]
201        & 6                         & \classS{}                 &                 & 3798679053239405824       & J11261592+0142258    & 201578138  & --         & 28.2       & 1.5        & 0.95                           & 11.1                           &           \\ 
           &                           &                           &                 & 3798679053239406080       & J11261688+0142237    & 201578098  & --         & 27.6       & 2.5        & 1.43                           & 12.5                           &           \\ 
\\[-0.6em]
220        & 6                         & \classS{}                 &                 & 2075134756418822272       & J19561853+4108112    & --         & 5905418    & 30.5       & 1.5        & 1.07                           & 12.3                           &           \\ 
           &                           &                           &                 & 2075134962577250176       & J19561703+4108289    & --         & 5905382    & 34.1       & 1.5        & 2.27                           & 15.6                           &           \\ 
\\[-0.6em]
225        & 4                         & \classS{}                 &                 & 3584374199647645056       & J12204839-0647402    & 228799745  & --         & 23.6       & 2.5        & 0.93                           & 12.2                           &           \\ 
           &                           &                           &                 & 3584374203942504448       & J12204894-0647371    & 228799765  & --         & 19.3       & 2.8        & 1.68                           & 14.0                           & Double    \\ 
\\[-0.6em]
246        & 4                         & \classS{}                 &                 & 3596051051789926528       & J12005429-0513184    & 201153090  & --         & 23.6       & 2.3        & 1.19                           & 12.6                           &           \\ 
           &                           &                           &                 & 3596051086149665280       & J12005483-0512490    & 201153439  & --         & 25.0       & 2.0        & 2.1                            & 15.0                           &           \\ 
\\[-0.6em]
292        & 4                         & \classE{}                 &                 & 2052430769006849792       & J19365340+3928181    & --         & 4375408    & 1.18       & 0.01       & 0.61                           & 10.0                           & Subgiant  \\ 
           &                           &                           &                 & 2052430769006846720       & J19365272+3928089    & --         & 4375393    & 14.4       & 0.1        & 0.8                            & 12.8                           &           \\ 
\\[-0.6em]
299        & 3                         & \classW{}                 &                 & 2053584770878226304       & --                   & --         & --         &  --        &  --        & -0.1                           & 17.9                           & White Dwarf\\ 
           &                           &                           &                 & 2053585565450552960       & J19284890+4100216    & --         & 5792108    & 13.6       & 0.1        & 0.98                           & 11.3                           &           \\ 
\ldots     & \ldots                    & \ldots                    & \ldots          & \ldots               & \ldots     & \ldots     & \ldots     & \ldots     & \ldots     & \ldots                         & \ldots                         & \ldots    \\ 
    \hline
\end{tabular}
    \tablefoot{
        The table lists the individual stars and groups them into the individual binaries. The stars listed are those shown in Fig.\,\ref{fig_age_bins} with two more examples of wide binaries that contain an evolved component (see Sect.\,\ref{sec_extrapolate}). 
        The ``Binary ID'' column refers to the enumeration we have introduced for our sample of \allbinaries{} wide binaries. The ``Age group'' column refers to the assignment performed in Sect.\,\ref{sec_analysis}. The ``System Class'' column refers to the classification from the same section (see table footnote (a) for details). The ``System Issue'' column lists the potential explanation for an inconsistent system, and the ``Notes on components'' column lists any peculiarities of a (component) star, such as signs of (close) binarity or an evolved state. The complete table, which has additional columns (e.g., astrometry), is available in electronic form.\\
    \tablefoottext{a}{The ``Class'' column contains the classification from Sect.\,\ref{sec_analysis} and is as follows: 
        \classS{} = \primary{} component and \secondary{} component on slow rotator sequence; 
        \classF{} = at least one of the components is a fast rotator; 
        \classC{} = \secondary{} component contradicts \primary{} component; 
        \classE{} = binary contains a (sub)giant; 
        \classW{} = binary contains a white dwarf; 
    }
    }
\end{sidewaystable*}

\end{appendix}
 
\end{document}